\documentclass[12pt,tabularx,amsmath,a4paper]{article}

\usepackage{setspace,graphics}
\usepackage[dvips]{epsfig} 
\usepackage{a4,amssymb,epsfig,array,cite,setspace}
\usepackage[hypertex]{hyperref}
\usepackage{slashed}

\newcounter{multieqs}

\newcommand{\be}{\begin{equation}}
\newcommand{\ee}{\end{equation}}
\newcommand{\eq}[1]{(\ref{#1})}

\def\nn{\nonumber}
\def\bea{\begin{eqnarray}}
\def\eea{\end{eqnarray}}
\def\obar{\overline}

\def\beqa{\begin{eqnarray}} 
\def\eeqa{\end{eqnarray}} 
\def\beq{\begin{equation}} 
\def\eeq{\end{equation}}

\def\Tr{{\rm Tr}}

\def\a{\alpha}

\def\d{\delta}  \def\D{\Delta}

 \def\L{\Lambda}

\def\cA{{\cal A}}  \def\cC{{\cal C}}
 \def\cE{{\cal E}} 
 \def\cH{{\cal H}} 
  
\def\cM{{\cal M}}

\def\R{{\mathbb R}}

\def\one{\mbox{1 \kern-.59em {\rm l}}}

\def\bit{\begin{itemize}}
\def\eit{\end{itemize}}

\def\({\left(}
\def\){\right)}

\def\d{\delta}\def\D{\Delta}

\def\uno{\mbox{1 \kern-.59em {\rm l}}}

\newcommand{\tr}{\mbox{tr}}

\def\bcomment#1{}

\newcommand{\lek}{\left[}
\newcommand{\rek}{\right]}
\newcommand{\lrk}{\left(}
\newcommand{\rrk}{\right)}

\newcommand{\bsp}{\begin{split}}
\newcommand{\esp}{\end{split}}
\newcommand{\phanti}{\phantom{i}}
\newcommand{\thetainv}{\theta^{-1}}

\renewcommand{\title}[1]{\vspace{10mm}\noindent{\Large{\bf #1}}\vspace{8mm}}
\newcommand{\authors}[1]{\noindent{\large #1}\vspace{5mm}}
\newcommand{\address}[1]{{\itshape #1\vspace{2mm}}}

\begin{document}

\begin{titlepage}

\begin{flushright}
UWTHPh-2008-08\\
\end{flushright}

\begin{center}
  
\title{Fermions and Emergent Noncommutative Gravity} \\
\vspace{1cm}

\authors{Daniela {\sc Klammer}${}^{1}$ and Harold {\sc Steinacker}${}^{2}$}

 \address{ Fakult\"at f\"ur Physik, Universit\"at Wien\\
 Boltzmanngasse 5, A-1090 Wien, Austria \\}

\footnotetext[1]{daniela.klammer@univie.ac.at}
\footnotetext[2]{harold.steinacker@univie.ac.at}

\vskip 2cm

\textbf{Abstract}

\vskip 3mm 

\begin{minipage}{14cm}%

Fermions coupled to Yang-Mills matrix models are studied from the point
of view of emergent gravity. We show that the simple matrix model
action provides an appropriate coupling for fermions to gravity, albeit
with a non-standard spin connection. 
Integrating out the fermions in a nontrivial geometrical background 
induces indeed the Einstein-Hilbert action and a dilaton-like term, 
at least for on-shell geometries. 
This explains and precisely reproduces the UV/IR mixing for fermions in 
noncommutative gauge theory, extending
recent results for scalar fields.
It also explains why some UV/IR mixing
remains even in supersymmetric models, except in the $N=4$ case.

\end{minipage}

\end{center}

\end{titlepage}

\setcounter{page}0
\thispagestyle{empty}

\begin{spacing}{.3}
{
\noindent\rule\textwidth{.1pt}            
   \tableofcontents
\vspace{.6cm}
\noindent\rule\textwidth{.1pt}
}
\end{spacing}


\section{Introduction}

Recently it was understood that gravity can arise effectively 
from non-commutative gauge theory 
\cite{Rivelles:2002ez,Yang:2006mn,Steinacker:2007dq}.
This mechanism is realized in certain matrix models of Yang-Mills
type, which are known to describe gauge theory on non-commutative
(NC) spaces. 
These models can be interpreted more naturally as describing dynamical
NC spaces, suggesting an intrinsic realization
of some sort of gravity theory. 
The mechanism of this ``emergent gravity'' is quite simple 
and intrinsically non-commutative \cite{Steinacker:2007dq}:
On a general background corresponding to a generic NC space, 
all fields in the model couple universally 
(up to possibly density factors) 
to an effective metric or frame associated 
with the background. 
The Einstein-Hilbert action is induced upon quantization.
Such a generic NC background can be equivalently interpreted
in terms of 
a $U(1)$ gauge field on a fixed Moyal-Weyl space $\R^4_\theta$.
This then provides an understanding of 
UV/IR mixing in the 
one-loop effective action of NC gauge theory, 
in terms of an induced gravity action. 

In the present paper, we include fermions in these matrix models,
and study their coupling to the emergent gravity. 
In particular, we extend the explanation of UV/IR mixing 
in terms of induced gravity  
\cite{Grosse:2008xr} to the case of fermions.
The matrix-model framework strongly suggests a simple
action for fermions, as realized e.g. in the string-theoretical 
IKKT  model. We study this fermionic action in detail on a 
generic NC background,
following the geometrical point of view 
of \cite{Steinacker:2007dq}. This leads to an effective 
semi-classical action for a fermion on a background
with metric $G_{ab}(y)$. 
The action is similar to the standard action for fermions
on a curved background, except that the spin-connection 
appears to be missing. It is nevertheless well-defined, due to 
the existence of an (unobservable) preferred frame, which is 
an intrinsic part of emergent NC gravity.

The main result of this paper is that in spite of this 
unusual feature, the simple fermionic action under consideration
is nevertheless reasonable on general backgrounds, 
and should be physically viable. In the point particle limit,
the fermions will follow the same trajectories as
for the conventional coupling to gravity, 
albeit with a different rotation of the spin along the trajectory. 
Furthermore, we determine 
the effective gravitational action obtained by integrating out
the fermions in our framework. This leads indeed to the expected 
Einstein-Hilbert action for the effective metric, 
with an extra term for a density factor, 
and another term which vanishes on-shell. 
We conclude that the fermionic
action considered here is the ``correct'' one
for emergent gravity, suitable for 
a physically realistic theory of 
gravity. 

For a consistent quantization of emergent gravity, 
a supersymmetric extension of the model appears to be
necessary. We point out that the induced gravitational action 
will be finite if and only if the model enjoys $N=4$ supersymmetry
at the Planck scale. This explains in particular the
persistence of UV/IR mixing in NC gauge theory in models with
less supersymmetry. Such a supersymmetric extension is
realized by the IKKT model, which is known to admit noncommutative
backgrounds; see e.g. 
\cite{Ishibashi:1996xs,Aoki:1999vr,Ishibashi:2000hh,Kitazawa:2005ih}
and references therein
for related work including (indirect) evidence for gravity on such 
backgrounds.

This paper is organized as follows. In Section 2, we 
review the basic aspects of emergent gravity, and 
write down the actions under consideration.
In Section 3, we review the mechanism of induced gravity,
and set up the form of the
appropriate Seeley-de Witt coefficients corresponding to the
fermionic action. They are the ingredients which 
provide the induced effective gravity action, and they
are modified in comparison with the commutative case.
These coefficients are computed in 
Section 4, which is the core of this paper. We also provide 
an exact expression for the Ricci scalar in terms of the 
effective metric of emergent gravity; this should
be useful for other considerations as well. 
In Section 5, we consider the same problem from the
point of view of NC gauge theory, and 
rewrite the obtained gravitational action
in terms of $U(1)$ gauge fields on Moyal-Weyl space.
The one-loop effective action for this NC gauge theory
is carefully computed in the appropriate IR limit, 
which provides a precise matching with the gravitational 
action obtained before. 
We conclude with a short discussion of supersymmetry
and an outlook.

\section{Matrix models and effective geometry}
\label{sec:metric}

As a starting point of emergent NC gravity, we consider the 
semi-classical geometry arising from the matrix model with action 
\be
S_{YM} = - Tr [Y^a,Y^b] [Y^{a'},Y^{b'}] g_{a a'} g_{b b'}\, .
\label{YM-action-1}
\ee
Here $Y^a, \,\, a=1,2,3,4$ are hermitian matrices 
or operators acting on some Hilbert space $\cH$, which
constitute the dynamical objects of the model. The 
model also contains a constant background metric, which is 
\be
g_{a a'} = \delta_{a a'} \quad \mbox{or}\quad g_{a a'} = \eta_{a a'} 
\label{background-metric}
\ee
in the Euclidean  resp.  Minkowski case. 
The commutator of 2 matrices is denoted as 
\be
[Y^a,Y^b] = i \theta^{ab}\, ,
\ee
defining $\theta^{ab} \in L(\cH)$ as an antihermitian 
operator-valued matrix. 
We focus here on configurations where the $Y^a$ 
(which need not be solutions of 
the equation of motion) can be interpreted as quantization 
of coordinate functions $y^a$ on 
a Poisson manifold $(\cM,\theta^{ab}(y))$ with general
Poisson structure $\theta^{ab}(y)$. This defines the geometrical 
background under consideration, and conversely
any Poisson manifold provides (locally) after quantization a possible
background $Y^a$. 
More formally, this means that there is an isomorphism of vector spaces
\be
\begin{array}{ccl}
\cC(\cM) &\to& \cA\,\subset \, L(\cH)\, \\
 f(y) &\mapsto& \hat f(Y) \\
i\{f,g\} &\mapsto& [\hat f,\hat g] + O(\theta^2)
\end{array}
\label{map} 
\ee
where $\cC(\cM)$ denotes the space of functions on $\cM$,
and $\cA$ is the algebra generated by $Y^a$,
interpreted as quantized algebra of functions. 
Thus $\theta^{ab}$ reduces in a semi-classical limit 
to a classical Poisson structure $\theta^{ab}(y)$. 
Observe that the $Y^a$ correspond to preferred
coordinate functions $y^a$ on $\cM$, and $g_{ab} = \d_{ab}$
resp. $g_{ab} = \eta_{ab}$ defines a flat ``background'' metric, 
which is 
constant in the preferred coordinates $y^a$. 
Indices throughout this work will always refer to these
preferred coordinates\footnote{this is the reason why many of the 
formulas in this paper are written in a non-covariant way.
A covariant formulation is possible but will not be given here.},
and only some (final) formulas will be manifestly covariant.

To simplify things, we restrict ourselves to the
purely geometrical or  ``irreducible'' case in this paper,
i.e. we assume that $\cA$ is in some sense dense in $L(\cH)$. 
This means that any matrix in $L(\cH)$  
can be considered as a function of $Y^a$ resp. $y^a$.
From the gauge theory point of view in section 
\ref{change-of-variables}, it 
means that we restrict ourselves to the $U(1)$ case;
this is most interesting here since the UV/IR mixing 
is restricted to the trace-$U(1)$ sector. 
For the nonabelian case see \cite{Steinacker:2007dq}.

Besides the 
preferred coordinates $y^a$, the noncommutative background 
provides in the semi-classical limit a preferred frame
\be
e^a = -i [Y^a,.] = \theta^{ab} \partial_b 
\label{frame}
\ee
given in terms of the antisymmetric 
Poisson tensor, which we assume to be non-degenerate. 
This formula is only valid for the 
preferred coordinates $y^a$, and does not admit local
Lorentz transformations. The effective metric arising from the 
matrix model turns out to be  \cite{Steinacker:2007dq}
\be
G^{ab}(y) = \theta^{ac}(y) \theta^{b d}(y)\, g_{cd} \, ,
\label{effective-metric}
\ee
which has the correct tensor structure and
is associated to the above frame. 
$G^{ab}(y)$ is indeed 
the effective gravitational metric in the matrix model
(up to a rescaling discussed below),
because it enters the kinetic terms for matter fields through
$[Y^a,\Psi] \sim i\theta^{ab}(y) \frac{\partial}{\partial y^b} \Psi$;
this will be seen explicitly below.
Note that $G^{ab}(y)$ is not flat in general. 
Hence the background $\cM$ naturally 
acquires a metric structure $(\cM,\theta^{ab}(y),G^{ab}(y))$, 
which is determined by the Poisson structure and the 
constant background metric $g_{ab}$.
While this is rather obvious for scalar fields,
it turns out that essentially the same $G^{ab}$ 
also couples to nonabelian gauge fields
\cite{Steinacker:2007dq}, and to fermions as we will show here.

An infinitesimal version of the metric \eq{effective-metric} 
was observed already in \cite{Rivelles:2002ez}. 
The frame \eq{frame} was also pointed out in \cite{Yang:2006mn}, 
as well as a metric of type \eq{effective-metric} 
for the self-dual case. 
There is also some overlap with the ideas in \cite{Madore:2000aq}.
For further related work see e.g. 
\cite{Muthukumar:2004wj,Banerjee:2004rs,Fatollahi:2005ri}.

\paragraph{Scalars.}

We first review the case of scalar fields 
i.e. hermitian matrices $\Phi$ coupled to 
the matrix model \eq{YM-action-1}. 
The only possibility 
to write down kinetic terms for matter fields is through commutators 
$[Y^a,\Phi] \sim i\theta^{ab}(y) \frac{\partial}{\partial y^b} \Phi$,
and one is lead to the action
\bea
S[\Phi] &=& -(2\pi)^2\, \Tr\, g_{aa'} [Y^a,\Phi][Y^{a'},\Phi]  \nn\\
&\sim& \int d^4 y\, \rho(y)\, G^{ab}(y)\,\frac{\partial}{\partial y^a}\Phi(y) 
\frac{\partial}{\partial y^b} \Phi(y) .
\label{scalar-action-0}
\eea
Here and throughout this paper, $\sim$ 
indicates the leading contribution in a semi-classical 
expansion in powers of $\theta^{ab}$. This
involves the symplectic measure
\be
\rho(y) =  (\det\theta^{ab}(y))^{-1/2} = |G_{ab}(y)|^{1/4}  
\equiv e^{-\sigma}
\quad (\,\equiv \L_{NC}^4(y)\,)
\ee
on $(\cM,\theta^{ab}(y))$,
which can be naturally interpreted as non-commutative 
scale $\L_{NC}^4$; here
$|G_{ab}(y)| \equiv |\det G_{ab}(y)|$.
After appropriate rescaling of $G^{ab}(y)$, this can be 
rewritten in covariant form 
\be
S[\Phi] = \int d^4 y\, \tilde G^{ab}(y) \partial_{y^a}\Phi \partial_{y^b}\Phi 
\label{scalar-action-geom}
\ee
with the effective metric 
\bea
\tilde G^{ab} &=& |G_{ab}|^{1/4}\, G^{ab} = \rho(y)\, G^{ab} \nn\\
|\tilde G^{ab}| &=& 1\, 
\label{metric-unimod}
\eea
which is unimodular in the preferred $y^a$ coordinates.
Recall that these are characterized by the background
metric being constant, $g_{ab} = \d_{ab}$ resp. $g_{ab} = \eta_{ab}$.

\paragraph{Fermions.}

Then the most obvious 
(perhaps the only reasonable) action
for a spinor which can be written down in the matrix 
model framework\footnote{In particular, 
fermions should also be in the adjoint, otherwise they
cannot acquire a kinetic term. This does not rule out its applicability
in particle physics, see e.g. \cite{Steinacker:2007ay}.} is
\be
S = (2\pi)^2\, \Tr  \obar\Psi  \gamma_a [Y^a,\Psi]
 \,\, \sim\,\, \int d^4 y\, \rho(y)\, \obar\Psi i \gamma_a \theta^{ab}(y)
\partial_b \Psi
 \label{fermionic-action-geom}
\ee
ignoring possible nonabelian gauge
fields here to simplify the notation. This is written for the case of
Minkowski signature, the Euclidean version involves the obvious
replacement $\bar\Psi \to \Psi^\dagger$.
This defines the (matrix) Dirac operator
\be
\not \!\! D \,\Psi = \gamma_a [Y^a,\Psi]\, 
\,\, \sim\,\, i \gamma_a \theta^{ab}(y)\partial_b \Psi.
\label{Diracop-matrix}
\ee 
We can compare this with the 
standard covariant derivative for spinors (see e.g. \cite{weinberg})
\be
\not \!\! D_{\rm comm} \, \Psi 
= i\gamma^a\, e_a^\mu \(\partial_\mu + \Sigma_{bc}\, \omega^{bc}_\mu\)\Psi 
\label{covar-spinor}
\ee
where  
\be
\omega^{ab}_\mu = i\frac 12\, e^{a\nu} \(\nabla_\mu e^b_{\nu}\)
\label{spinconnection-2}
\ee
is the spin connection, and
\be
\Sigma_{ab} = \frac i4 [\gamma_a,\gamma_b] \, .
\ee
Comparing \eq{Diracop-matrix} with \eq{covar-spinor},
we observe again that in the 
geometry defined by \eq{effective-metric},
\be
e^\mu_b(y) := \theta^{\mu c}(y) g_{cb} 
\label{vielbein}
\ee
plays the role of a preferred vielbein. However this must
be used with great care, because the distinction between 
the coordinate index $\mu$ and the Lorentz index $a$ is lost
in the special ``gauge'' inherent in \eq{vielbein}.

One notices immediately that 
the spin connection appears to be missing in
\eq{fermionic-action-geom}, 
which seems very strange 
at first. One of the main messages of this work is that 
in spite of this strange feature, the action \eq{fermionic-action-geom}
is a good action for a fermion propagating in the geometry
defined by $\tilde G_{ab}$. 
Recall that the spin connection determines how the spinors are rotated 
under parallel transport along a trajectory.
However, the spin-connection $\omega^{ab}_\mu$ can 
always be eliminated 
(via parallel-transport resp. a suitable gauge choice)
along an open trajectory. Then the conventional
kinetic term \eq{covar-spinor} 
boils down to \eq{fermionic-action-geom}.
Therefore in the point-particle limit,
the trajectory of a fermion with action
\eq{fermionic-action-geom} 
will follow properly the geodesics 
of the metric\footnote{for massless particles, the geodesics of 
$\tilde G_{ab}$ coincide with those of $G_{ab}$.
Masses should be generated spontaneously, which is not considered
here.} $\tilde G_{ab}$, 
albeit with a different rotation
of the spin. Furthermore, we will show in detail that the
induced gravitational action obtained by integrating out the 
fermion in \eq{fermionic-action-geom}
indeed induces the expected Einstein-Hilbert term
$\int d^4 y\,  R[\tilde G]\, \L^2$ at least
for ``on-shell geometries'', albeit with an 
unusual numerical coefficient and an extra term 
depending on $\sigma$.
All this shows that \eq{fermionic-action-geom} defines a
reasonable action for fermions in the background defined by 
$\tilde G_{ab}$.

In particular, the transport along a closed curve determines
the holonomy in a gravitational field, and the missing spin connection
in the above action strongly suggests that holonomies here
will be different than in General Relativity. 
More generally, the gravitational 
spin rotation for a free-falling fermion might provide
a nice signature for or against the emergent gravity framework.

\paragraph{Equations of motion.}

So far we considered arbitrary background configurations $Y^a$ as long
as they admit a geometric interpretation.
The equations of motion  derived from the action \eq{YM-action-1} 
are
\be \label{eom}
[Y^a,[Y^{a'},Y^b]]\, g_{a a'} = 0 \, , 
\ee
which in the semi-classical limit amount to 
\be
\theta^{ma}\partial_m \theta^{nb} g_{ab}=0
\qquad \mbox{or}\qquad \tilde G^{ac}\partial_c \theta^{-1}_{cd} =0\, .
\label{eom-theta}
\ee
They
select on-shell geometries among all possible backgrounds, such as 
the Moyal-Weyl quantum plane \eq{Moyal-Weyl}. 
In the present geometric form they amount to Ricci-flat
spaces \cite{Rivelles:2002ez,Steinacker:2007dq} 
at least in the linearized case. 
However since we are interested in the quantization
here, we will need general off-shell configurations below.
 The equations \eq{eom-theta}
are only valid in the preferred $y^a$ coordinates as discussed above, and the same 
applies to most computations below.
They can be cast in covariant form, which will not be done here.

\section{Quantization and induced gravity}

We are interested in the quantization of our matrix model coupled to fermions.
In principle, the quantization is defined in terms of a (``path'') integral
over all matrices $Y^a$ and $\Psi$. 
In four dimensions, we can only perform perturbative computations
for the ``gauge sector'' $Y^a$, while the fermions can be integrated
out formally in terms of a determinant. Let us focus here on the
effective action $\Gamma_{\Psi}$ obtained by integrating out the fermionic fields,
\be
e^{-\Gamma_{\Psi}} = \int d\Psi\,d\bar{\Psi}\; e^{-S[\Psi]}, 
\ee
which for non-interacting fermions is given by 
\be \label{trlog2}
\Gamma_{\Psi}=- \frac 12\Tr \log \not \!\! D^2.
\ee
Later, we will consider an alternative interpretation as
Laplacian of the fermionic fields on $\R^4_{\obar \theta}$ coupled to an
adjoint $U(1)$ gauge field. 
In Feynman diagram language, \eq{trlog2} will then
amount to the sum of all one-loop diagrams with arbitrary
numbers of external $A$-lines.

In~\cite{Grosse:2008xr},  the analogous
computation of the effective action $\Gamma_{\Phi}$ for scalar fields
was carried out 
in two different ways. On the one hand $\Gamma_{\Phi}$ was evaluated
from the geometric 
point of view via induced Einstein-Hilbert action. On the other hand the action 
\eq{scalar-action-0} was regarded as NC gauge theory by means of
covariant coordinates, 
where the effective action for the NC gauge fields was determined by
integrating out the scalar fields.   
The obtained effective actions were shown to agree in the IR regime as
expected. 
As a consequence UV/IR mixing of NC gauge theory can be interpreted as an effect of gravity.

In this paper, we generalize the 
results of~\cite{Grosse:2008xr} to the case of fermions. 
The effective action $\Gamma_{\Psi}$ is evaluated 
in two different ways: We first compute $\Gamma_{\Psi}$  
as induced gravity action, using its semi-classical geometrical 
form. This is then compared with the 
one-loop effective action of NC $\mathfrak{u}(1)$ gauge theory.
As expected, we find complete agreement in a suitable IR regime.

\paragraph{Square of the Dirac operator and induced action.}

Starting from the action 
\be
S = (2\pi)^2\, \Tr  \Psi^\dagger  \gamma_a [Y^a,\Psi] 
\ee
we want to study
\bea
 e^{-\Gamma_{\Psi}}&=& \int d\Psi d\obar \Psi\, 
e^{ -(2\pi)^2\, Tr  \Psi^\dagger  \gamma_a [Y^a,\Psi]}
  \, \cong\, \det ( \gamma_a [Y^a,.]) \nn\\
 &=& \exp\(\ln\det(\not \!\! D)\)
= \exp\(\frac 12 \log\det(\not \!\! D^2)\) \nn\\
&=& \exp\(\frac 12 \Tr \log(\not \!\! D^2)\) 
\eea
at one loop, considering the Euclidean case for the sake of rigor.
The square of the  Dirac operator takes the following form
\bea
\not \!\! D^2 \Psi &=& \gamma_a\gamma_b [Y^a,[Y^b,\Psi]] \nn\\
&=& -\gamma_a\gamma_b \theta^{ac} \partial_c (\theta^{bd} \partial_d \Psi) \nn\\
&=&  - G^{cd} \partial_c\partial_d  \Psi
  -  a^d\partial_d \Psi,
\label{D-square-1}
\eea
with
\be
a^d = \gamma_a\gamma_b \theta^{ma}\partial_m \theta^{db} 
=-2i\,\Sigma_{ab}\theta^{ac}\partial_c\theta^{bd} + g_{ab}\theta^{ac}\partial_c\theta^{bd}.
\label{a-def}
\ee
See Appendix C for a comparison of this term with 
the commutative case.
$\not \!\! D^2$ defines the quadratic form
\bea\label{S-square}
S_{\rm square} &=& (2\pi)^2\, \Tr \Psi^\dagger\not \!\! D^2 \Psi 
\sim \int d^4y\, \rho(y) \Psi^\dagger\not \!\! D^2 \Psi \nn\\
&=&  \int d^4y\, |G_{ab}|^{1/4} \Psi^\dagger\not \!\! D^2 \Psi,
\eea
which is very similar to the scalar action.
In terms of the  metric $\tilde G_{ab}$ \eq{metric-unimod}
with $|\tilde G_{ab}| =1$, $S_{\rm square}$ 
can be written in covariant form
\be
S_{\rm square} = \int d^4y\, \sqrt{|\tilde G|}\,\,  \obar\Psi\,
\widetilde{\not \!\!  D^2}\Psi ,
\ee
in terms of the rescaled squared Dirac operator 
\bea
\widetilde{\not \!\! D^2} \Psi &=& -\left(
\tilde G^{cd} \partial_c\partial_d  \Psi
  + e^{-\sigma}\,a^d \partial_d \Psi  \right).
\label{D-squared}
\eea
We now compute the effective action using
\bea
\frac 12 \Tr (\log \widetilde{\not \!\! D^2} - \log \widetilde{\not \!\! D_0^2}) 
&=& -\frac 12 \Tr \int_{0}^\infty d\a \frac 1\a 
\(e^{-\a  \,\widetilde{\not \! D^2}} - e^{-\a  \,\widetilde{\not \! D_0^2}}\) \nn\\
&\equiv&\,\, -\frac 12 \Tr\int_{0}^{\infty} \frac{d\a}{\a}\,
\Big(e^{-\a  \, \widetilde{\not \! D^2}} - e^{-\a \,\widetilde{\not \! D_0^2}
}\Big)\, e^{- \frac 1{2\a \tilde \L^2}} ,
\label{heatkernel-expand}
\eea
where $\widetilde{\L}^2$ denotes the cutoff\footnote{We write $2\L^2$
  instead of $\L^2$
in \eq{heatkernel-expand} in order to be consistent with the
cutoff for $\frac 12 \D^2$ for scalar fields used in \cite{Grosse:2008xr}, which
is implicit in $\Gamma_\Phi$ as given below.} for 
$\frac 12\widetilde{\not \!\! D^2}$, regularizing the divergence 
for small $\alpha$. Now we can apply the heat kernel expansion
\be
\Tr\, e^{-\alpha \;\widetilde{\not \! D^2}} 
=
\sum_{n\geq 0}\,\alpha^{\frac{n-4}{2}}\, \int_{\cM}d^4 y \;a_n\left(y,\widetilde{\not \!\! D^2}\right)
\ee
where the Seeley-de Witt coefficients $a_n (y,\widetilde{\not \!\! D^2})$ are given by~\cite{Gilkey:1995mj}
\bea
a_0(y) &=& \frac{1}{16\pi^2}\,\tr\,\one,\nn\\
a_2(y) &=& \frac{1}{16\pi^2}\tr\left(\frac{R[\widetilde{G}]}{6}\;\one + \cE \right) \nn\\
\cE &=& - \widetilde{G}^{mn}\left(
\partial_m \Omega_n + \Omega_m\Omega_n -\widetilde{\Gamma}^k_{mn}\Omega_k,
\right)  \label{E-def}\\
\Omega_m &=& \frac{1}{2}\widetilde{G}_{mn}\left(e^{-\sigma}a^n +
  \widetilde{\Gamma}^n \right) \label{Omega-def}
\eea
and $\tr$ denotes the trace over the spinorial matrices.
The effective action is therefore
\bea \label{induced gravity action}
\Gamma_{\Psi} &=& \frac 1{16\pi^2}\, \int d^4 y \,\( 2\, \tr(\one)\,\widetilde{\L}^4 
 + \tr \left(\frac{R[\tilde G]}{6}\; \one +\cE\right)\,
 \widetilde{\L}^2 + O(\log \tilde \L) \)\, ,
 \label{S-oneloop-fermions}
\eea
where $\tr (\one) =4$ for a Dirac fermion.
Everything is expressed in terms of the unimodular metric
$\tilde{G}_{ab}$, 
which can be written in terms of $G_{ab}$ using
\bea
R[\widetilde{G}]&=& \rho(y)\left( R[G]+3\Delta_G \sigma -\frac 32 G^{ab}\partial_a \sigma \partial_b \sigma \right) \nn\\
\Delta_G \sigma &=& -(G^{ab}\partial_a \partial_b \sigma - \Gamma^c \partial_c \sigma) \nn\\
\Gamma^a &=& G^{bc}\Gamma^a_{bc} \nn\\
e^{-\sigma(y)}&=&\rho(y)=\lrk\det G_{ab}\rrk^{1/4} \nn\\
\widetilde{\Gamma}^a&=& \widetilde{G}^{cd}\widetilde{\Gamma}^a_{cd}
 \, =\,e^{-\sigma}\Gamma^a - e^{-\sigma}\lrk\partial_b\sigma\rrk G^{ba}.
\eea
The actual computation of $\tr\,\cE$ is given in Section \ref{Tr E}.
Note the relative minus sign of the various terms in the 
effective action $\Gamma_{\Psi}$ compared with the 
induced action due to a scalar field \cite{Grosse:2008xr}, 
\be
\Gamma_{\Phi}=\frac{1}{16\,\pi^2}\int d^4 y
\left(-2\widetilde{\Lambda}^4 
-\frac{1}{6} R[\widetilde{G}] \widetilde{\Lambda}^2 
+O(\log \widetilde{\Lambda})
\right).
\label{Gamma-phi}
\ee
hence
\be
\Gamma_\Psi + 4\, \Gamma_\Phi 
\,\, =\,\,  \frac{1}{16\,\pi^2}\int d^4 y \,\tr\, \cE \, \widetilde{\Lambda}^2 \,.
\label{induced-susy-E}
\ee
This expresses the cancellation of the induced actions due to 
fermions and bosons, apart from the $\cE$ term.
For the standard coupling of Dirac fermions to gravity
on commutative spaces, one has~\cite{Vassilevich:2003xt}
\be\label{cE-commutative}
\tr\,\cE_{\rm comm} = - R  
\ee
which originates from an additional constant term $-\frac 14 R$ in 
$\not \!\! D^2_{\rm comm}$ (Lichnerowicz's formula).
In our case, $\cE$ turns out to be somewhat modified
due to the vanishing spin connection, nevertheless it 
contains the appropriate curvature scalar plus an additional 
term \eq{trE-result}.
This will be discussed further in Section \ref{susy-cancellations}.

\section{Computation of the Seeley-DeWitt coefficient}
\label{sec:SdW}

In this section we determine the second Seeley-de Witt 
coefficient for \eq{D-squared}. 
We first obtain an exact, compact result for 
the Ricci scalar $R[\widetilde{G}]$ 
expressed in terms of the Poisson tensor $\theta^{mn}(y)$.
This may be useful for other purposes as well. 
We then compute $\tr\,\cE$ in terms of $\theta^{mn}(y)$, which
turns out to be closely related to $R[\widetilde{G}]$
as desired.

\subsection{Ricci scalar in terms of $\theta^{mn}$}
The curvature is given as usual by
\be
{R_{abc}}^d = \partial_b \Gamma^{d}_{ac} - \partial_a\Gamma^{d}_{bc}
 + \Gamma^{e}_{ac}\Gamma^{d}_{eb} - \Gamma^{e}_{bc}\Gamma^{d}_{ea}\, .
\ee
The Ricci scalar is then
\bea
R= G^{ac}\,R_{abc}^b= G^{ac}\left(\partial_b \Gamma_{ac}^b - \partial_a\Gamma^b_{bc}
+\Gamma^e_{ac}\Gamma^b_{eb}-\Gamma^e_{bc}\Gamma^b_{ea}\right).
\eea
In terms of the metric and its derivatives $R$ is given by
\bea
R&=&\lrk\partial_b G^{bd}\rrk G^{ca}\lrk\partial_a G_{cd}\rrk 
+G^{ab}G^{cd}\partial_b\partial_d G_{ac} 
-G^{mn}G^{pq}\partial_p\partial_q G_{mn} \nonumber\\
&\quad&
-\lrk\partial_b G^{bd}\rrk G^{mn}\lrk\partial_d G_{mn}\rrk
-\frac{3}{4}G^{pq}\lrk\partial_p G^{mn}\rrk\lrk\partial_q G_{mn}\rrk \nn\\
&\quad&
+\frac{1}{2}G^{np}\lrk\partial_p G^{ac}\rrk\lrk\partial_c G_{na}\rrk
-\frac{1}{4} G^{pq}G^{mn}\lrk\partial_p G_{mn} \rrk G^{kl}\lrk\partial_q G_{kl}\rrk.
\eea
We use the explicit formula for the metric tensor
\be
G^{mn}(y)=\theta^{ma}(y)\theta^{nb}(y) g_{ab}
\ee
to express $R$ in terms of $\theta$ (see Appendix A for details)
\bea
R
&=&
- 2\lrk\partial_a \theta^{ap}\rrk G^{bc}\lrk\partial_c \thetainv_{bp}\rrk 
-\lrk\partial_a \theta^{ap}\rrk\lrk\partial_b \theta^{bq}\rrk g_{pq} \nn\\
&\quad&
+2 G^{mp}G^{nq}\thetainv_{ma}\partial_p\partial_q\thetainv_{nb} g^{ab}
+ \frac{1}{2} G^{bc}\lrk\partial_c\thetainv_{ap}\rrk G^{ad}\lrk\partial_d \thetainv_{bq}\rrk g^{pq}
\nn\\
&\quad&
+2\theta^{mn}G^{pq}\partial_p\partial_q \thetainv_{mn} 
-\frac{1}{2} G^{mn}G^{pq}\lrk \partial_p \thetainv_{ma}\rrk\lrk\partial_q\thetainv_{nb}\rrk g^{ab} 
\nn\\
&\quad&
+4\theta^{bc}\lrk\partial_b\theta^{da}\rrk\lrk\partial_d\sigma\rrk g_{ca}
+\frac{3}{2} G^{pq}\lrk\partial_p \theta^{mn}\rrk\lrk\partial_q \thetainv_{mn}\rrk
\nn\\
&\quad&
-G^{mp}\lrk\partial_p \theta^{nq}\rrk\lrk\partial_n \thetainv_{mq}\rrk \quad
-\frac{1}{2}\lrk\partial_n\theta^{cq}\rrk\lrk\partial_c\theta^{np}\rrk g_{pq}.
\eea
This equation holds in fact for any vielbein
using the identification
\begin{eqnarray}
\theta^{\mu m}g_{mb}=e^\mu_{\phanti b} \qquad  
\theta^{-1}_{\mu m}g^{mb}=-e_\mu^{\phanti b}
\end{eqnarray}
since we have not exploited any property of $\theta^{mn}$ yet. 
However, by making use of the Jacobi identity,
\bea
\partial_a \theta^{-1}_{bc}+\partial_c \theta^{-1}_{ab}+\partial_b \theta^{-1}_{ca}=0\\
\partial_a \theta^{pq}=-\lrk \partial_c \theta^{-1}_{am}\rrk\lrk \theta^{mp}\theta^{cq}-\theta^{mq}\theta^{cp}\rrk,
\eea
several terms appearing in the computation of $\tr\,\cE$ and 
$R[\widetilde{G}]$ are equivalent\footnote{By means of these 
relations one can also check that the action (\ref{S-square}) 
is indeed hermitian.}:
\bea \label{eq: jacobi-relations}
G^{mk}\lrk\partial_k\theta^{na}\rrk\lrk\partial_n\thetainv_{ma}\rrk
&=&
\frac{1}{2}G^{pq}\lrk\partial_p\theta^{mn}\rrk\lrk\partial_q\thetainv_{mn}\rrk\nn\\
\lrk\theta^{mn}\partial_q\thetainv_{mn}\rrk\theta^{qa} G^{pk}
\lrk\partial_k\thetainv_{pa}\rrk 
&=&
2\lrk\partial_m\theta^{ma}\rrk G^{nk}\lrk\partial_k\thetainv_{na}\rrk\nn\\
\lrk\theta^{mn}\partial_q\thetainv_{mn}\rrk\theta^{qa}\lrk\partial_p\theta^{pb}\rrk g_{ab}
&=&
2\lrk\partial_m\theta^{ma}\rrk\lrk\partial_n\theta^{nb}\rrk g_{ab}\nn\\
\theta^{mn}G^{pq}\partial_p\partial_q\thetainv_{mn}&=&-2 G^{mp}G^{nq}\thetainv_{ma}\partial_p\partial_q\thetainv_{nb} g^{ab}\nn\\
\theta^{ma}\partial_m\partial_n\theta^{nb} g_{ab}
&=&
G^{mn}\partial_m\partial_n\sigma +\theta^{ma}\lrk\partial_m\theta^{nb}\rrk\lrk\partial_n\sigma\rrk g_{ab}
\nn\\
&=&
\frac{1}{2}\theta^{mn}G^{pq}\partial_p\partial_q \thetainv_{mn}
+\frac{1}{2}G^{pq}\lrk\partial_p\theta^{mn}\rrk\lrk\partial_q \thetainv_{mn}\rrk \nn\\
&\quad& + \lrk\partial_p\theta^{pa}\rrk G^{qk}\lrk\partial_k \thetainv_{qa}\rrk
\eea
Thus in our framework the Ricci scalar associated to 
$G_{ab}$ takes the form
\bea
R[G]&=& \theta^{mn}G^{pq}\partial_p\partial_q \thetainv_{mn} + G^{pq}\lrk\partial_p\theta^{mn}\rrk\lrk\partial_q\thetainv_{mn}\rrk \nn\\
&\quad&
+2\lrk\partial_m\theta^{ma}\rrk G^{nk}\lrk\partial_k\thetainv_{na}\rrk - G^{mn}\lrk\partial_m\sigma\rrk\lrk\partial_n\sigma\rrk \nn\\
&\quad&
+\frac{1}{2}G^{mk}\lrk\partial_k\thetainv_{na}\rrk G^{nl}\lrk\partial_l\thetainv_{mb}\rrk g^{ab}
-\frac{1}{2}G^{mn}G^{pq}\lrk\partial_p\thetainv_{ma}\rrk\lrk\partial_q\thetainv_{nb}\rrk g^{ab} \nn\\
&\quad&
-\frac{1}{2}\lrk\partial_m\theta^{na}\rrk\lrk\partial_n\theta^{mb}\rrk g_{ab}.
\eea
Evaluating also
\bea
3\Delta_G\sigma-\frac{3}{2}G^{mn}\partial_m \sigma \partial_n\sigma &=&
-\frac{3}{2} G^{mn}\lrk\partial_m \theta^{pq}\rrk\lrk\partial_n\thetainv_{pq}\rrk 
-\frac{3}{2}\theta^{mn} G^{pq}\partial_p\partial_q \thetainv_{mn} \nn\\
&\quad&
+\frac{3}{2} G^{mn}\partial_m\sigma\partial_n\sigma - 3 \lrk \partial_m\theta^{ma}\rrk G^{nk}\lrk\partial_k \thetainv_{na}\rrk \nn\\
\eea
gives 
\bea
R[\widetilde{G}]&=e^{-\sigma}&\left(
R\left[G\right]+3\Delta_G \sigma -\frac{3}{2} G^{ab}\partial_a \sigma \partial_b \sigma_b
\right)\nn\\
&=
e^{-\sigma}&\Big[
\;
-\frac{1}{2}\theta^{mn}G^{pq}\partial_p\partial_q \thetainv_{mn} -\frac{1}{2} G^{pq}\lrk\partial_p\theta^{mn}\rrk\lrk\partial_q\thetainv_{mn}\rrk \nn\\
&\quad&
-\lrk\partial_m\theta^{ma}\rrk G^{nk}\lrk\partial_k\thetainv_{na}\rrk +\frac{1}{2} G^{mn}\lrk\partial_m\sigma\rrk\lrk\partial_n\sigma\rrk \nn\\
&\quad&
+\frac{1}{2}G^{mk}\lrk\partial_k\thetainv_{na}\rrk G^{nl}\lrk\partial_l\thetainv_{mb}\rrk g^{ab}
-\frac{1}{2}G^{mn}G^{pq}\lrk\partial_p\thetainv_{ma}\rrk\lrk\partial_q\thetainv_{nb}\rrk g^{ab} \nn\\
&\quad&
-\frac{1}{2}\lrk\partial_m\theta^{na}\rrk\lrk\partial_n\theta^{mb}\rrk g_{ab}
\Big] .
\label{R-exact}
\eea
Via partial integration, the number of independent terms 
can be further reduced: 
\bea
\int d^4 y\;e^{-\sigma}\theta^{ma}\partial_m\partial_n\theta^{nb} g_{ab}&=&0, \nn\\
\int d^4 y\;e^{-\sigma}\lrk\partial_m\theta^{na}\rrk\lrk\partial_n\theta^{mb}\rrk g_{ab} &=& -\int d^4 y\; \frac{e^{-\sigma}}{2}\left(
\theta^{mn}G^{pq}\partial_p\partial_q\thetainv_{mn}
+G^{pq}\lrk\partial_p\theta^{mn}\rrk\lrk\partial_q\thetainv_{mn}\rrk\right), \nn\\
\int d^4 y\;e^{-\sigma}\lrk\partial_p\theta^{pa}\rrk G^{qk}\lrk\partial_k \thetainv_{qa}\rrk
&=& 
\int d^4 y\;e^{-\sigma}\lrk\partial_m\theta^{na}\rrk\lrk\partial_n\theta^{mb}\rrk g_{ab}\, .
\eea
This yields the following compact form for the Ricci scalar  
\bea
\int d^4 y\, R[\widetilde{G}]\widetilde{\Lambda}^2&=&
\int d^4 y\; e^{-\sigma} \Big\{
\frac{1}{2}G^{mk}\lrk\partial_k\thetainv_{na}\rrk G^{nl}\lrk\partial_l\thetainv_{mb}\rrk g^{ab}
-\frac{1}{2}G^{mn}G^{pq}\lrk\partial_p\thetainv_{ma}\rrk\lrk\partial_q\thetainv_{nb}\rrk g^{ab} \nn\\
&\quad&
-\frac{1}{2}\lrk \partial_p \theta^{pa}\rrk G^{qk}\lrk\partial_k \thetainv_{qa}\rrk
+\frac{1}{2} G^{mn}\lrk\partial_m \sigma\rrk\lrk\partial_n \sigma\rrk 
\Big\}\widetilde{\Lambda}^2 \, ,
\label{R-action-explicit}
\eea
which is indeed what we need because $\det\tilde G_{ab} =1$.
However, one has to be careful when using partial integration. 
We regard here the cutoff $\widetilde{\Lambda}^2$ as a 
constant effective cutoff for $\Delta_{\widetilde{G}}$,
independent of $y^a$. In Section \ref{change-of-variables}, 
we will relate $\widetilde{\Lambda}^2$ with the effective cutoff 
$\Lambda$ for $\Delta_A$, which is the Laplacian of NC gauge theory. 
This corresponds to a position-dependent
$\widetilde{\Lambda}^2$ as given in 
\eq{cutoff relation}. In that case, one must either 
include additional terms from partial integration, or use
\eq{R-exact} as we will do.


\subsection{Evaluation of $\tr\,\cE$} \label{Tr E}

We need to evaluate 
\bea
\tr \cE&=&-\tr\widetilde{G}^{ab}\left(\partial_a \Omega_b + \Omega_a\Omega_b - \widetilde{\Gamma}^r_{ab}\Omega_r\right), 
\eea
where
\bea \label{omega-ausruck}
\Omega_m&=&\frac{1}{2}\widetilde{G}_{mn}\left(\widetilde{a}^n+\tilde{\Gamma}^n\right)\nn\\
&=&
\frac{1}{2}\left(
G_{mn}\gamma_a\gamma_b\theta^{pa}\left(\partial_p \theta^{nb}\right)
-G_{mn}\lrk\partial_p G^{pn}\rrk +\partial_m \sigma
\right)
\eea
and $\tilde a^n = e^{-\sigma}\, a^n$.
For the computation of $\tr\,\cE$ we use the relations given in 
Appendix A and the Jacobi
identity (\ref{eq: jacobi-relations}). Recalling
\bea
\tr \gamma^a \gamma^b &=& 4 g^{ab},\nn\\
\tr \gamma^a \gamma^b \gamma^c \gamma^d &=& 
4\left(g^{ab}g^{cd}-g^{ac}g^{bd}+g^{ad}g^{bc} \right)\, ,\nn\\
\eea 
we find for the individual parts of $\tr\,\cE$:
\bea
\tr\,\widetilde{G}^{mn}\partial_m\Omega_n &=&
2e^{-\sigma}\Big\{
G^{mn}\lrk\partial_m G_{np}\rrk g_{ab} \theta^{qa}\lrk\partial_q\theta^{pb}\rrk \nn\\
&\quad&
+\lrk\partial_m\theta^{qa}\rrk\lrk\partial_q\theta^{mb}\rrk g_{ab} 
+\theta^{ma}\partial_m\partial_n\theta^{nb}g_{ab} \nn\\
&\quad&
-G^{mn}\lrk\partial_m G_{np}\rrk\lrk\partial_q G^{qp}\rrk 
-\partial_m\partial_n G^{mn}
+G^{mn}\partial_m\partial_n \sigma 
\Big\}\nn\\
&=&
2e^{-\sigma}\Big\{
\lrk\partial_m\theta^{ma}\rrk G^{nk}\lrk\partial_k\thetainv_{na}\rrk - \theta^{ma}\partial_m\partial_n\theta^{nb} g_{ab}
 \nn\\
&\quad&
+\frac{1}{2}G^{pq}\lrk\partial_p\theta^{mn}\rrk\lrk\partial_q\thetainv_{mn}\rrk +\frac{1}{2}\theta^{mn}G^{pq}\partial_p\partial_q \thetainv_{mn}
\Big\} \nn\\
&=&0  \\[3ex]
\tr\,\widetilde{G}^{mn}\Omega_m\Omega_n &=&
e^{-\sigma}\Big\{
\left(g_{ab}g_{cd}-g_{ac}g_{bd}+g_{ad}g_{bc}\right)
G_{nk}\theta^{qa}\lrk\partial_q\theta^{nb}\rrk\theta^{lc}\lrk\partial_l\theta^{kd}\rrk \nn\\
&\quad&
-2\lrk\partial_q G^{qn}\rrk G_{nk}g_{cd}\theta^{lc}\lrk\partial_l\theta^{kd}\rrk 
+2\theta^{qa}\lrk\partial_q\theta^{nb}\rrk\lrk\partial_n\sigma \rrk g_{ab} \nn\\
&\quad&
-2\lrk\partial_l G^{lk}\rrk\lrk\partial_k\sigma\rrk +\lrk\partial_q G^{qn}\rrk\lrk\partial_l G^{lk}\rrk G_{nk}
+G^{mn}\lrk\partial_m\sigma\rrk\lrk\partial_n \sigma\rrk 
\Big\}\nn\\
&=&
e^{-\sigma}\Big\{
- G^{kl}G^{mn}\lrk\partial_k\thetainv_{ma}\rrk\lrk\partial_l\thetainv_{nb}\rrk g^{ab} 
+\,G^{mk}\lrk\partial_k\thetainv_{na}\rrk G^{nl}\lrk\partial_l\thetainv_{mb}\rrk g^{ab}
\Big\} \nn
\\[3ex]
\tr\,\Omega_m \widetilde{\Gamma}^m
&=&
e^{-\sigma}\Tr\left(\Omega_m \Gamma^m - \Omega_m G^mn\partial_n\sigma\right)\nn\\
&=&e^{-\sigma}\Tr\left(-\Omega_m\lrk\partial_n G^{nm}\rrk 
+ \Omega_m G^{mn}\partial_n\sigma \right)
\nn\\
&=& 0
\eea
Hence we obtain
\bea \label{TrE}
\tr\,\cE &=&
e^{-\sigma} \Big\{
G^{kl}G^{mn}\lrk\partial_k\thetainv_{ma}\rrk\lrk\partial_l\thetainv_{nb}\rrk g^{ab} 
-G^{mk}\lrk\partial_k\thetainv_{na}\rrk G^{nl}\lrk\partial_l\thetainv_{mb}\rrk g^{ab}
\Big\}.
\eea
Comparing with \eq{R-action-explicit}
for $\widetilde{\Lambda}^2$ regarded as constant cutoff 
of $\Delta_{\widetilde{G}}$, we can write this as
\bea
\int d^4 y\;\tr\,\cE &=& \int d^4 y\;
\(-2\,R[\tilde G]
-\lrk \partial_p \theta^{pa}\rrk G^{qk}\lrk\partial_k \thetainv_{qa}\rrk
+ G^{mn}\lrk\partial_m \sigma\rrk\lrk\partial_n \sigma\rrk \) \nn\\
&\stackrel{\mathrm{eom}}{=}& 
\int d^4 y\;\(-2R[\tilde G]+G^{mn}\partial_m\sigma\partial_n\sigma\),
\label{trE-result}
\eea
assuming on-shell geometries \eq{eom-theta} in the last line.
This formula applies for Dirac fermions, and 
with an additional factor $\frac 12$ for Weyl fermions.
It is remarkable that $\tr\,\cE$ is essentially given 
by the appropriate 
curvature scalar $R[\tilde G]$, and a contribution from the 
dilaton-like scaling factor $\rho = e^{-\sigma}$. This is a very
reasonable modification of the standard result 
\eq{cE-commutative}, as desired.


\section{Relation with gauge theory 
on $\mathbb{R}_{\theta}^4$}
\label{change-of-variables}

We now 
want to interpret the fermionic action \eq{fermionic-action-geom} as 
action for a Dirac fermion on the Moyal-Weyl quantum
plane $\R^4_\theta$
coupled to a $U(1)$ gauge field in the adjoint. 
This point of view
is obtained by writing the general covariant coordinate
resp. matrix $Y^a$ as
\be
Y^a = X^{a} + \cA^a\, .
\label{cov-coord-1}
\ee
Here $X^{a}$ are generators of the Moyal-Weyl quantum plane, which 
satisfy
\be
[X^a,X^b] = i \bar\theta^{ab}\, ,
\label{Moyal-Weyl}
\ee
where $\bar \theta^{ab}$ is a constant antisymmetric tensor. 
These are particular
 solutions of the equations of motion \eq{eom}.
The effective geometry \eq{effective-metric} 
for the Moyal-Weyl plane is indeed flat, given by
\bea
\bar g^{ab} &=& \bar\theta^{ac}\,\bar\theta^{bd} g_{cd}\, \nn\\
\tilde g^{ab} &=& \bar \rho\, \bar g^{ab} ,
\qquad \det \tilde g^{ab} =1  \nn\\
\bar\rho &=&  (\det\bar\theta^{ab})^{-1/2}
= |\bar g_{ab}|^{1/4} \equiv \L_{NC}^4 \, .
\label{effective-metric-bar}
\eea
Consider now the
change of variables
\be
\cA^a(x) = -\bar\theta^{ab} A_b(x)
\label{A-naive}
\ee
where $A_a$ are hermitian matrices, interpreted  as smooth functions 
on $\R^4_{\bar \theta}$. 
 Thus we can write
\bea
\lek Y^a, f \rek=\lek X^a + \cA^a, f\rek = i \bar{\theta}^{ab}\lrk \frac{\partial }{\partial x^b}f + i\lek A_b, f\rek \rrk 
\equiv i\bar{\theta}^{ab} D_b f,
\eea
giving for the quadratic form \eq{S-square}
\bea
S_{square}&=& (2\pi)^2\,  \Tr\; \Psi^\dagger \gamma_a\gamma_b\lek Y^a,
\lek Y^b, \Psi\rek \rek \nn\\
&=&-\int d^4x \,\bar{\rho} \,\Psi^\dagger\;\gamma_a \gamma_b 
\bar{\theta}^{am}\bar{\theta}^{bn}D_m D_n \Psi \nn\\
&=&\int d^4 x\, \Psi^\dagger \,\widetilde{\not \!\! D^2}_A \Psi\, . 
\eea
This is an exact expression on $\R^4_\theta$,
where
\be
\widetilde{\not \!\! D^2}_A 
= - \bar{\rho} \,\gamma_a\gamma_b\bar{\theta}^{am}\bar{\theta}^{bn}D_m D_n  \, 
= -  \,\tilde\gamma^m\tilde\gamma^n\, D_m D_n  \, ,
\ee
and
\be
\tilde \gamma^a = (\det \bar g_{ab})^{\frac 18}\,\gamma_b\, \bar\theta^{ba},   \qquad
\{\tilde \gamma^a,\tilde\gamma^b\} =  2 \,\tilde g^{ab}\, .
\label{tilde-gamma}
\ee
We now want to rewrite the geometrical results of Section \ref{sec:SdW}
in terms of gauge theory on $\R^4_\theta$ in $x$-coordinates. To do this, 
note that most formulas of Section \ref{sec:SdW}
are not generally covariant, but only valid in the preferred
$y$-coordinates defined by the matrix models where
$g_{ab} = \d_{ab}$ resp. $g_{ab} = \eta_{ab}$.
\eq{cov-coord-1} defines the 
leading-order relation between $y$ and $x$ coordinates, 
\be
y^a = x^a - \bar{\theta}^{ab} \bar{A}_b + O(\theta^2)\, ,
\ee
with Jacobian
\bea
\left\vert \frac{\partial y^a}{\partial x^b}\right\vert &=& 
1-\bar{\theta}^{ac}\frac{\partial A_c}{\partial x^a} + O(\theta^2) \nn\\
&=&1 - \frac{1}{2}\bar{\theta}^{mn}\bar{F}_{mn} + O(\theta^2).
\label{jacobian}
\eea
See ~\cite{Grosse:2008xr} for details of this change of variables.
In order to avoid confusion we will denote all $x$-dependent tensors 
with a bar, and we write
\be
\partial_a = \frac{\partial}{\partial y^a}, \qquad \bar{\partial}_a=\frac{\partial}{\partial x^a}.
\ee
The Poisson tensor can be written 
in terms of the $\mathfrak{u}(1)$ field strength as 
\be
i \theta^{ab}(y)=\lek Y^a, Y^b\rek 
= i\bar{\theta}^{ab} - i\bar{\theta}^{ac}\bar{\theta}^{bd}\bar{F}_{cd},
\ee
where $\bar{F}_{cd}$ is a rank two tensor in $x$ coordinates
 on $\mathbb{R}_{\theta}^4$.
This amounts to 
\bea
\thetainv_{ab}=\bar{\theta}^{-1}_{ab}-\bar{F}_{ab}
\eea
at leading order.
We also need the metric \eq{effective-metric} in $x$-coordinates, 
 \bea
G^{ab}&=&\lrk \bar{\theta}^{ac}-\bar{\theta}^{ai}\bar{\theta}^{cj}\bar{F}_{ij}\rrk
 \lrk \bar{\theta}^{bd}-\bar{\theta}^{be}\bar{\theta}^{df}\bar{F}_{ef}\rrk g_{cd} \nn\\
&=&
\bar{g}^{ac}\left(
\delta^b_c + \bar{F}_{cd}\bar{\theta}^{db}+\bar{g}_{ce}\bar{\theta}^{ef}\bar{F}_{fd}\bar{g}^{db}
+\bar{g}_{cd}\bar{\theta}^{de}\bar{F}_{ef}\bar{g}^{fg}\bar{F}_{gh}\bar{\theta}^{hb}
\right)\nn\\
&\equiv& \bar{g}^{ac}\lrk \delta^b_c + X_c^b \rrk.
\eea
To compute the determinant, we use
\bea
\det\lrk 1 + X_{ij}\rrk =
1 + tr X +\frac 12 \lrk \lrk tr X\rrk^2 - tr\lrk X^2\rrk \rrk + O(X^3).
\eea
Then
\bea
tr X &=& -2\bar{F}_{mn}\bar{\theta}^{mn} - \bar{\theta}^{em}\bar{g}_{mn}\bar{\theta}^{fn}\bar{F}_{eh}\bar{g}^{hg}\bar{F}_{gf}
\eea
and
\bea
e^{\sigma}&=& |G^{ab}|^{1/4}=|\bar{g}^{ab}|^{1/4}
\left(
1 - \frac 12 \bar{\theta}^{mn}\bar{F}_{mn} + O(\bar{\theta}^3)
\right). 
\eea
By a straightforward application of the relations given above one can now write down the second Seeley-de Witt coefficient in $x$-coordinates. 
$R[\tilde G]$ expressed in $x$-coordinates agrees with eq. (78) 
in~\cite{Grosse:2008xr}, as it should be. 
$\tr\,\cE$ in $x$-coordinates is given as 
\bea
\tr\cE &=&
|\bar g_{ab}|^{1/4}\left(
-\frac{1}{4}\bar{\theta}^{mn}\bar{F}_{mn}\bar{\partial}^a\bar{\partial}_a\bar{\theta}^{pq}\bar{F}_{pq}
-\frac{1}{2}\bar{g}^{ac}\bar{g}^{bd}\bar{F}_{ab}\bar{\partial}^2 \bar{F}_{cd}
\right)\nn\\
&\quad&
+\frac{1}{4}|\bar g_{ab}|^{1/4}\bar{\theta}^{mn}\bar{F}_{mn}\bar{\partial}^a\bar{\partial}_a\bar{\theta}^{pq}\bar{F}_{pq}
\nn\\
&=&
-|\bar g_{ab}|^{1/4}\frac{1}{2}\bar{g}^{ac}\bar{g}^{bd}\bar{F}_{ab}\bar{\partial}^2 \bar{F}_{cd},
\eea
where
\be
\bar{\partial}^2=\bar{\partial}_a\bar{\partial}_b g^{ab}.
\ee
In Appendix B this expansion 
is given explicitly for the individual terms. 
We have omitted $O(A)$ terms from both $R[\widetilde{G}]$ and 
$\tr\,\cE$, which are total derivatives and do not contribute 
to the effective action. 
In this way, we find for the one-loop induced action 
\bea
\Gamma_{\Psi}&=&\int d^4 y \left(
a_0 \widetilde{\Lambda}^4 + a_2 \widetilde{\Lambda}^2 + O\left(\log \widetilde{\Lambda}\right)\right)
\nn\\
&=&-4 \Gamma_{\Phi}-\frac{1}{16\,\pi^2}\int d^4y \frac{\rho(y)}{2}\bar{g}^{ac}\bar{g}^{bd}\bar{F}_{ab}\bar{\partial}^2 \bar{F}_{cd} \,
\widetilde{\Lambda}^2.
\eea
Finally, there is a nontrivial relation between the
cutoff $\widetilde{\Lambda}$ of the geometrical action
and the cutoff $\L$ of the $\mathfrak{u}(1)$ gauge theory,
which follows from the identity
\bea
S_{\rm square}=\Tr\; \Psi^\dagger\gamma_a\gamma_b \lek Y^a,\lek Y^b,\Psi\rek \rek 
= \int d^4y\,\Psi^\dagger\widetilde{\not \!\! D^2}_{\widetilde{G}}\Psi   
= \int d^4 y \frac{\rho(y)}{\bar{\rho}}\Psi^\dagger\widetilde{\not \!\! D^2}_A \Psi.
\eea
For the Laplacians this means
\be
\widetilde{\not \!\! D^2}_{\widetilde{G}} =\frac{\rho(y)}{\bar{\rho}}\widetilde{\not \!\! D^2}_A.
\ee
Since we implement the cutoffs using the Schwinger parameterization \eq{heatkernel-expand},
they are related as follows (cf. \cite{Grosse:2008xr})
\bea \label{cutoff relation}
\widetilde{\Lambda}^2=\frac{\rho(y)}{\bar{\rho}}\Lambda^2.
\eea
This makes sense provided $\rho(y)/\bar{\rho}$ varies only on large scales respectively small momenta $p\ll \Lambda$, which is our 
working assumption. Together with \eq{jacobian}, we 
obtain as a final result for the geometric
one-loop effective action 
expressed in terms of gauge theory on $\R^4_\theta$
\bea
\Gamma_{\Psi}&=&-4 \Gamma_{\Phi}-\int d^4x \bar{\rho} \frac{\Lambda^2}{2}\bar{g}^{ac}\bar{g}^{bd}\bar F_{ab}\bar{\partial}^2 \bar F_{cd} \nn\\
&=&-4 \Gamma_{\Phi}
+\int \frac{d^4 p}{(2\pi)^4}  \widetilde{g}^{ac}\widetilde{g}^{bd}\bar F_{ab}(p) \bar F_{cd}(-p)\frac{p^2}{\Lambda_{NC}^{4}}\frac{\Lambda^2}{2}
\label{S-eff-geom-expand}
\eea
where $p^2 = p_i p_j g^{ij}$. This agrees 
precisely with the one-loop computation in the
gauge theory point of view  obtained below.
Note that the last term corresponds to $\tr\,\cE$
in \eq{induced-susy-E}.

\subsection{Comparison with UV/IR mixing}
\label{sec:UV-IR-mixing}

In this section, we compare the geometrical form of the one-loop
effective action obtained in the
previous section with the one-loop effective action obtained from
the gauge theory point of view. The result is of course the same, 
which provides not only a nontrivial check for our geometrical 
interpretation, but also sheds new light on the conditions
to which extent the semi-classical analysis of the previous section is
valid. This generalizes the results of \cite{Grosse:2008xr} 
to the fermionic case.
We find as expected that the UV/IR mixing terms obtained by
integrating out the fermions are given by the 
induced geometrical resp. 
gravitational action \eq{S-oneloop-fermions}, in a suitable IR regime.
In particular, we need an explicit, physical momentum cutoff $\L$.

Using the variables and conventions of the previous section,
the action \eq{fermionic-action-geom} can be exactly 
rewritten as $U(1)$ gauge theory on $\R^4_\theta$,
which in the Euclidean case takes the form
\bea
S[\Psi] &=& (2\pi)^2\,  \Tr \Psi^\dagger \gamma_a [Y^a,\Psi]  \nn\\
&=& \int  d^4 x\, \tilde\Psi^\dagger
i\tilde \gamma^a (\bar\partial_a\tilde\Psi +i g[A_a, \tilde\Psi])
\eea
We introduce an explicit coupling constant
$g$, and define a rescaled fermionic field 
\be
\tilde \Psi = |\bar g_{ab}|^{\frac 1{16}}\,\Psi
\label{psi-tilde}
\ee
in order to
obtain the properly normalized effective metric $\tilde g^{ab}$;
we will omit 
the tilde on $\Psi$ henceforth. Recall also that
only $U(1)$ gauge fields are considered here, 
because only those correspond to the nontrivial geometry
considered in the previous section. 

We need the $O(A^2)$ contribution to the one-loop effective action
obtained by integrating out the fermionic field $\Psi$.
While this  computation has been discussed several times in the 
literature 
\cite{Minwalla:1999px,Hayakawa:1999zf,Matusis:2000jf,VanRaamsdonk:2001jd,Khoze:2000sy}, 
the known results are 
not accurate enough for our purpose, i.e. 
in the regime $p^2, \L^2<\L_{NC}^2$ 
where the semiclassical geometry is 
expected to make sense. 
We need to analyze carefully the  IR regime of
the well-known effective cutoff $\L_{eff}(p)$ \eq{lambda-eff} 
for non-planar graphs as $p \to 0$, keeping $\L$ fixed. 
In this regime the non-planar diagrams 
almost coincide with the planar diagrams, and the leading IR corrections
due to the non-planar diagrams correspond 
to the induced geometrical terms in \eq{S-oneloop-fermions}.
This has not been considered in 
previous attempts to explain UV/IR mixing, e.g. in terms of
exchange of closed string modes \cite{Armoni:2001uw,Sarkar:2005jw}.

\begin{figure}[h]
\begin{center}
\includegraphics[scale=0.7]{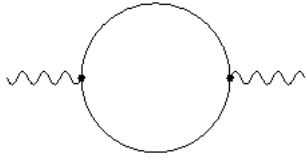}
\end{center}
\caption{fermionic one-loop diagram}
\label{fig:1}
\end{figure}

\begin{figure}[h]
\begin{center}
\includegraphics[scale=0.4]{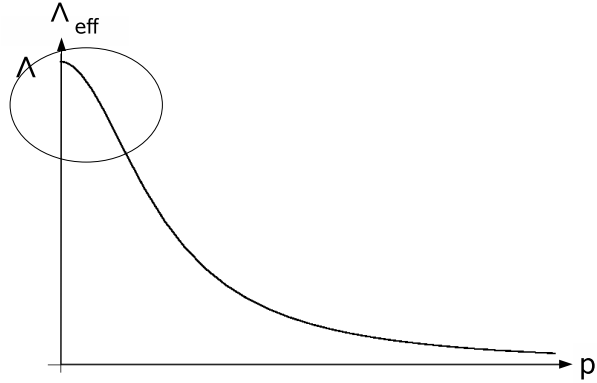}
\end{center}
\caption{relevant IR regime of $\L_{eff}(p)$}
\label{fig:L-eff}
\end{figure}

To proceed one can either square the Dirac operator as in 
\cite{Khoze:2000sy}, or use directly the fermionic Feynman rules. 
We choose the latter approach here, and consider the Feynman 
diagram in figure \ref{fig:1} corresponding to
\bea
\Gamma_{\Psi} &=& - \frac 12 \Tr \log \Delta_0 
-\frac{g^2}2 \left<\int d^4 x\, \bar\rho\, \bar \Psi \tilde\gamma^a [A_a, \Psi] 
\int d^4 y\, \bar\rho\, \bar \Psi \tilde\gamma^b [A_b, \Psi]\right>  \nn\\
&=& -\frac 12 \Tr \log \Delta_0 +\Gamma_{\Psi}(A) .
\eea
The minus sign in front is due to the fermionic loop.
This gives 
\bea
\Gamma_{\Psi} &=& - 4 g^2\int \frac{d^4 p}{(2\pi)^4}\, 
A_{a'}(p) A_{b'}(-p)\,\tilde g^{a'a} \tilde g^{b'b}\, \int \frac{d^4 k}{(2\pi)^4}\,
\frac{2k_a k_b + k_a p_b + p_a k_b - \tilde g_{ab} k(k+p)}
{(k\cdot k)((k+p)\cdot(k+p))}\, \nn\\
&& \Big(e^{-i k_i \theta^{ij} p_j}-1\Big)
\label{Gamma-fermi}
\eea
which is quite close to the bosonic case, using the notation
\bea
k\cdot k &\equiv& k_i k_j\, \tilde g^{ij} \, .  \nn\\
k^2 &\equiv& k_i\, k_j g^{ij}  \, .
\label{norm-notation}
\eea
The momentum integrals are understood to be regularized by a 
cutoff $\L$, implemented via a Schwinger parameter as in \cite{Grosse:2008xr}.
To evaluate this loop integral, we rewrite it in a 
different way as in \cite{Khoze:2000sy}
\bea
&&  -\int \frac{d^4 k}{(2\pi)^4}\,
\frac{4k_a k_b + 2 k_a p_b + 2 p_a k_b - 2\tilde g_{ab} k(k+p)}
{(k\cdot k)((k+p)\cdot(k+p))}\big(e^{-i k_i \theta^{ij} p_j}-1\big)\,
 \nn\\
&=& - \int \frac{d^4 k}{(2\pi)^4}\, 
 \Big(\frac{(2k_a + p_a)(2 k_b + p_b) - (p_a p_b - \tilde g_{ab} p\cdot p)}
   {(k\cdot k)((k+p)\cdot(k+p))} \nn\\
&&\quad  - \tilde g_{ab} 
\Big(\frac 1{k\cdot k} + \frac 1{(k+p)\cdot(k+p)}\Big) \Big)
 \big(e^{-i k_i \theta^{ij} p_j}-1\big)\nn\\
&=& - \int \frac{d^4 k}{(2\pi)^4}\, 
 \Big(\frac{(2k_a + p_a)(2 k_b + p_b)}{(k\cdot k)((k+p)\cdot(k+p))} 
 - 2 \frac{\tilde g_{ab}}{k\cdot k}\Big)
  \big(e^{-i k_i \theta^{ij} p_j}-1\big) \nn\\
&&  +  (p_a p_b - \tilde g_{ab} p\cdot p) \int \frac{d^4 k}{(2\pi)^4}\,
\frac 1{(k\cdot k)((k+p)\cdot(k+p))} 
  \big(e^{-i k_i \theta^{ij} p_j}-1\big)
\label{Gamma-fermi-2}
\eea
where we replaced $\frac 1{(k+p)\cdot(k+p)}$ by $\frac 1{k\cdot k}$ 
under the integral
(which does not make a difference in the regularization used here).
Now the first term is precisely the induced action $\Gamma_\Phi$
obtained by integrating out a scalar field $\Phi$ 
\cite{Grosse:2008xr}, 
which is known to be gauge invariant. 
The second term is logarithmic and manifestly gauge-invariant.
Therefore 
\bea
\Gamma_{\Psi} &=& - 4\, \Gamma_{\Phi} 
+ g^2 n_f\, \int \frac{d^4 p}{(2\pi)^4}\, 
A_{a'}(p) A_{b'}(-p) \,\,\tilde g^{a'a} \tilde g^{b'b}\,
 (p_a p_b - \tilde g_{ab} p\cdot p) \,\nn\\
&& \int \frac{d^4 k}{(2\pi)^4}\, \frac 1{(k\cdot k)((k+p)\cdot(k+p))}  
\(e^{-i k_i \theta^{ij} p_j}-1\) \nn\\
&=& - 4\, \Gamma_{\Phi} 
- g^2 n_f\,\int \frac{d^4 p}{(2\pi)^4}\, 
A_{a'}(p) A_{b'}(-p)\,\tilde g^{a'a} \tilde g^{b'b}\,
 (p_a p_b - \tilde g_{ab} p\cdot p) \,\nn\\
&& \frac 1{8 \pi^2}\, \int_0^1 dz\, 
\Big(K_0(2\sqrt{\frac{z(1-z)p\cdot p}{\L^2}})
 - K_0(2\sqrt{\frac{z(1-z)p\cdot p}{\L_{eff}^2}}) \Big) \,,
\label{Gamma-fermi-3}
\eea
for Dirac fermions, where
\bea
\L_{eff}^2 &=& { 1 \over 1/\Lambda^2 + \frac 14 \frac{p^2}{\L_{NC}^4}} 
= \L_{eff}^2(p) 
\label{lambda-eff}
\eea
is the ``effective'' cutoff for non-planar graphs, and $\L_{NC}$ is
defined in \eq{effective-metric-bar}.
For the standard evaluation of the $k-$ integration see e.g. 
\cite{Grosse:2008xr}.
To proceed we consider the IR regime  
\be
\frac{p^2 \Lambda^2}{\L_{NC}^4} < 1 \, ,
\label{IR-regime}
\ee
see figure \ref{fig:L-eff}.
Then both $\L$ and $\L_{eff}$ are large, and we can 
use the asymptotic expansions 
\be
K_0\Big(2 \sqrt{\frac{m^2}{\L^2}}\Big)  
\,=\,  -  \(\gamma + \log(\sqrt{\frac{m^2}{\L^2}})\) 
\, + O\Big(\frac{m^2}{\L^2}\log(\frac{\L}{m})\Big) 
\label{bessel-expand}
\ee
which gives
\bea
\Gamma_{\Psi} + 4\, \Gamma_{\Phi}
&\sim& \,\frac{ g^2 n_f}{16\pi^2} \int \frac{d^4 p}{(2\pi)^4}\, 
A_{a'}(p) A_{b'}(-p) \,\tilde g^{a'a} \tilde g^{b'b}\,
 (p_a p_b - \tilde g_{ab} p\cdot p) 
\, \log\Big(\frac{\L_{eff}^2}{\L^2}\Big)   \nn\\
&=&  -  \frac 12\,\frac{ g^2 n_f}{16\pi^2} \int \frac{d^4 p}{(2\pi)^4}\, 
 \bar F_{ab} \bar F_{a'b'} \tilde g^{a'a} \tilde g^{b'b}
\, \log\Big(\frac{\L_{eff}^2}{\L^2}\Big) \, .
\label{ind-action-gauge-fermion}
\eea
The only approximation here is the expansion \eq{bessel-expand}
of the Bessel functions in \eq{Gamma-fermi-3}.
$\Gamma_{\Phi}$ is the 1-loop effective action for a 
(hermitian) scalar
field as computed in \cite{Grosse:2008xr},
\bea
\Gamma_{\Phi} &=& - \frac{g^2}2\frac{1}{16 \pi^2}\,\int \frac{d^4 p}{(2\pi)^4}\, \, 
 \Big(-\frac 1{6} \bar F_{ab}(p)  \bar F_{a'b'}(-p) 
\tilde g^{a'a} \tilde g^{b'b}
 \,\log(\frac{\L^2}{\L_{eff}^2}) \nn\\
&&\qquad + \frac 14 (\theta \bar F(p)) (\theta \bar F(-p))
 \Big(\L_{eff}^4 - \frac 16 p\cdot p\, \L_{eff}^2 
 +\frac{(p\cdot p)^2}{1800}\, 
(47-30\log({\textstyle\frac{p\cdot p}{\L_{eff}^2}})) 
 \Big)\Big) \, . \nn\\
\label{ind-action-gauge-boson}
\eea
These expressions are valid in the IR regime \eq{IR-regime}
$p\, \L  < \L_{NC}^2$ corresponding to  ``mild'' UV/IR mixing.
This is the same condition which was obtained 
for the bosonic case \cite{Grosse:2008xr}.
We can then use the expansions
\bea
\L_{eff}^2 &=& \Lambda^2 - p^2 \frac{\Lambda^4 }{4\L_{NC}^4} +... \, ,
\label{lambda-eff-expand-1}\\
\L_{eff}^4 &=& \Lambda^4 - p^2 \frac{\Lambda^6 }{2\L_{NC}^4} +... \, ,
\label{lambda-eff-expand-2}\\
\log(\frac{\L^2}{\L_{eff}^2})&=& 
\frac 14 \frac{p^2 \L^2}{\L_{NC}^4} + ... 
\label{lambda-eff-expand-3}
\eea
which gives
\bea
\Gamma_{\Psi} + 4\, \Gamma_{\Phi} 
&\sim&  \, \frac 14  \,\frac{g^2}{16\pi^2} \int \frac{d^4 p}{(2\pi)^4}\, 
 \tilde g^{a'a} \tilde g^{b'b}\, \bar F_{ab}(p) \bar F_{a'b'}(-p)\,
 \frac{p^2\L^2}{\L_{NC}^4} \, ,  \nn\\
&=&  \, \frac 14  \,\frac{g^2}{16\pi^2} \int \frac{d^4 p}{(2\pi)^4}\, 
 \bar\rho^2\L^2  p^2 \,\bar g^{a'a}\bar g^{b'b}\, 
 \bar F_{ab}(p) \bar F_{a'b'}(-p)\, ,
\label{ind-action-gauge-boson-fermion}
\eea
where $p^2 = p_a p_b g^{ab}$. There are
obvious modifications due to the appropriate expansion of $\L_{eff}^2$
if one approaches the border of the IR regime \eq{IR-regime}.

To compare this with the geometrical results, we must take into
account the  different regularizations used in the heat-kernel 
expansion \eq{heatkernel-expand}
and in the above one-loop computation. 
It was shown in \cite{Grosse:2008xr} 
that these regularizations agree if we  replace $\L^2$ with $2 \L^2$
in the one-loop computation above\footnote{while this was strictly
  speaking established only for the bosonic case, the argument should 
extend to the fermionic case without difficulties.}. 
We then find  complete agreement with the result 
\eq{S-eff-geom-expand} obtained using
the geometrical point of view. 
Notice in particular that the induced gravitational action 
is nontrivial even in
the case of e.g. $N=1$ supersymmetry. This is now understood 
in terms of induced gravity, and full cancellation is obtained only
in the case of $N=4$ supersymmetry. This will be discussed below.

Finally, $\Gamma_\Psi$ and $\Gamma_\Phi$
 can be related directly to the geometrical 
induced action \eq{S-oneloop-fermions} in a 
more restricted IR regime, as in \cite{Grosse:2008xr}.
Assume first that

\underline{$\L \ll \L_{NC}$}.

Then the IR regime \eq{IR-regime} amounts to 
\be
p < \L_{NC},
\label{cutoff-simple}
\ee
which is very reasonable range of validity for the classical gravity
action. In this case,
\be
\L^6\frac{p^2}{\L_{NC}^4} = \frac{\L^4}{\L_{NC}^4} \L^2 p^2 
\,\,\ll\,\, \L^2  p^2 \,\,\sim\,\,  \L^2  p\cdot p 
\label{correction-cond}
\ee
so that we can replace
\be
\L_{eff}^4 - \frac 16 p\cdot p \L_{eff}^2 
\,\,\sim \,\,\L^4 - \frac 16p\cdot p \L^2\, .
\ee
Then the leading contribution to $\Gamma_\Phi$ is
\bea
\Gamma_{\Phi} &\sim&  -\frac{g^2}2\frac{1}{16 \pi^2}\,
\int \frac{d^4 p}{(2\pi)^4}\, 
 \Big(\frac{\L^4}4  (\theta \bar F(p)) (\theta \bar F(-p)) 
-\frac{\L^2}{24} \bar F_{ab}(p)\, \bar F_{a'b'}(-p) 
\frac{p^2}{\L_{NC}^4}
\tilde g^{a'a} \tilde g^{b'b} \nn\\
&& \qquad\qquad\quad - \frac{\L^2}{24}(\theta \bar F(p))
 (\theta \bar F(-p)) 
p\cdot p  \,\, +\, O(\log(\L))\,\, + \rm{finite}  \Big) \nn\\
&=& -\frac{g^2}2\frac{1}{16 \pi^2}\, \int d^4 x\, 
 \Big(\frac{\L^4}4 (\theta\bar F) (\theta\bar F)
- \bar \rho\,\frac{\L^2}{24} 
 \Big( p^2 \, \bar F_{ab}\, \bar F_{a'b'} \bar g^{a'a} \bar g^{b'b}
 + (p_a p_b \bar g^{ab}) (\theta \bar F)  (\theta \bar F) \Big) \nn\\
&& \qquad + O(\log(\L))\,\, + \rm{finite} \Big)\, ,
\label{Gamma-phi-F}
\eea
where again $p^2 = p_a p_b g^{ab}$. Taking into account again 
the appropriate replacement $\L^2 \to 2 \L^2$ corresponding
to the geometrical regularization in \eq{heatkernel-expand}
and setting $n_f=2$ for Dirac fermions, one finds 
as in \cite{Grosse:2008xr}
complete agreement between the above result for 
$\Gamma_\Psi$ with the result 
\eq{S-oneloop-fermions} obtained from
the geometrical point of view. 

Assume finally that the condition $\L \ll \L_{NC}$ is violated,
while maintaining the IR regime \eq{IR-regime}. 
Then there are additional terms in the effective action $\Gamma_\Phi$
beyond  \eq{Gamma-phi-F}.
They correspond to noncommutative corrections beyond the
semi-classical geometrical terms in \eq{Gamma-phi}; we refer to 
\cite{Grosse:2008xr} for some explicit results.

\paragraph{Gauge fields resp. gravitons}

The complete one-loop effective action for the gravitational sector
sector of the basic matrix model 
\eq{YM-action-1} contains contributions of scalar fields, fermions, 
and ``gauge fields''. The latter means trace- $U(1)$ components resp. 
gravitons inherent in $\tilde G_{ab}$, as 
well as possible nonabelian $SU(n)$ components
\cite{Steinacker:2007dq}.
Ignoring nonabelian fields for now, it remains to compute the 
contributions of the $U(1)$ gauge fields  in a loop
with 2 external $A$- legs.
This would lead to a similar contribution, denoted by
\be
e^{-\Gamma_A} = \int_{\rm one-loop} dA\, e^{-S}  \, .
\label{Gamma-A}
\ee
While $\Gamma_A$ is essentially straightforward 
to compute following e.g. \cite{Khoze:2000sy}, we can short-cut this
computation by taking advantage of the 
finiteness of the $N=4$ supersymmetric extension of the model. 
Note that the presence
of particular scalar interaction terms in the $N=4$ model is 
irrelevant for \eq{Gamma-A} at one loop. This leads to the result 
\eq{Gamma-A-explicit} given below.

\subsection{Cancellations and supersymmetry}
\label{susy-cancellations}

It is very interesting to compare the fermionic and the 
bosonic contribution to the gravitational action.
As is well-known \cite{Matusis:2000jf,Khoze:2000sy}, we note that the
fermionic contribution to the 
one-loop effective action in NC gauge theory
does not quite cancel the scalar contribution, due to 
\eq{ind-action-gauge-fermion}. 
This means that even in supersymmetric
cases some UV/IR mixing may remain. 
From the geometrical point of view, 
this terms corresponds to a gravitational action
$\tr\, \cE\, \tilde\L^2= - 2 R[\tilde G]\, \tilde\L^2 +
...$, so that the 
cutoff $\tilde\L^2$  should be interpreted as  
effective gravitational constant $\frac 1G$. 
This is completely analogous to the commutative case, where the
gravitational term \eq{cE-commutative} is induced.
The remaining UV/IR mixing
term cancels only in the case of $N=4$ supersymmetry.
We can therefore identify $\tilde\L$ as the scale of $N=4$ SUSY breaking
(assuming such a model), above which the model is finite. 
These observations strongly suggest that 
for the model to be well-defined at the quantum level,
$N=4$ SUSY
is required above the gravity scale i.e. the Planck scale.
This is  realized by the 
IKKT model \cite{Ishibashi:1996xs} on a NC background.

Let us therefore consider  an extension of the  matrix model 
\eq{YM-action-1} with $n_S$ scalar fields 
(= hermitian matrices $\phi^i$) and $n_\Psi$ Dirac fermions (hence $n_\Psi =
\frac 12$ for Weyl fermions).
The model with $N=4$ SUSY has $n_S = 6$ and 
$n_\Psi = 2$, in addition to the $U(1)$ gauge field $A_\mu$ 
resp. the graviton $h_{ab}$.
Taking its finiteness for granted, it follows that
\be
\Gamma_A = - 2\, \Gamma_\Psi - 6\, \Gamma_\Phi \, .
\label{Gamma-A-explicit}
\ee
This also holds for the  $SU(n)$ components of the
gauge fields in a nonabelian versions of this $N=4$ model; 
cf. \cite{Steinacker:2007dq}.
Using \eq{induced-susy-E}, this can be written explicitly as
\be
\Gamma_A = \frac{1}{16\,\pi^2}\int d^4 y \,
\left(-4\widetilde{\Lambda}^4 
- (\frac{1}{3} R[\widetilde{G}]+ 2\tr\, \cE) \widetilde{\Lambda}^2 \,
+O(\log \widetilde{\Lambda})
\right)\, .
\ee
This almost coincides with
the contribution of 2 scalars \eq{Gamma-phi}, except for  $\tr\, \cE$
which stands for \eq{trE-result}.
At this point, it is important to keep in mind that a term 
$\int d^4 y \,\sqrt{|\tilde G|}\,\widetilde{\Lambda}^4$ in the framework of 
emergent gravity does {\em not} amount to a large cosmological
constant, as discussed in 
\cite{Steinacker:2007dq,Grosse:2008xr}; see also 
\cite{Yang:2007as}. 
We only point out here that flat space is a solution even in the 
presence of this term. These issues will be discussed elsewhere 
in more detail.

\paragraph{$N=4$ supersymmetry transformations.}
The explicit form of the $N=4$ SUSY transformations 
 can be obtained as follows.
Consider the basic matrix model \eq{YM-action-1} 
in 10 dimensions with fermions, with action 
\be
S = - \Tr \(\frac 12 [Y^a,Y^b][Y^{a'},Y^{b'}]\eta_{aa'}\eta_{bb'} 
+ \bar \Psi \gamma_a [Y^a,\Psi] \),
\qquad a = 0,..., 9,
\label{action-susy}
\ee
where $\Psi$ is a 10-dimensional Majorana-Weyl spinor.
It is well known that this action
enjoys the 2 ``matrix'' supersymmetries \cite{Ishibashi:1996xs} 
\bea
\delta^{(1)} \Psi &=& \frac i2 [Y^a,Y^b] \gamma_{ab} \epsilon, \qquad 
\delta^{(1)} Y^a = i \bar \epsilon \gamma^a \Psi, \nn\\
\delta^{(2)} \Psi &=& \xi, \qquad \qquad  \qquad\quad \delta^{(2)} Y^a = 0
\label{susy-trafo-1}
\eea
where $\gamma_{ab} =\frac 12 [\gamma_a,\gamma_b]$,
and $\epsilon,\xi$ are Grassmann-valued spinors.
To recover spacetime supersymmetry, 
we split the matrices into 4 + 6 dimensions as
\be
Y^a = (Y^\mu,\phi^i), \qquad \mu = 0,...,3, 
\,\,\, i=1, ..., 6.
\label{extradim-splitting}
\ee
Then the 4-dimensional Moyal-Weyl quantum plane $\R^4_\theta$
is a (BPS) solution of the
generalized matrix equations of motion, embedded as
\bea
Y^\mu &=& \bar X^\mu, \qquad \mu= 0,...,3, \nn\\
\phi^i &=& 0\, .
\eea
All previous geometrical
considerations can be generalized,
except that the matrix model now contains scalar fields $\phi^i(y)$.
Even though we could consider a general metric 
$\tilde G_{ab}$ as above,
let us for simplicity 
only discuss the point of view of $U(1)$ gauge theory
on  $\R^4_\theta$.
If we set 
$\xi = \frac 12 \bar\theta^{\mu\nu} \gamma_{\mu\nu} \epsilon$
following \cite{Ishibashi:1996xs} 
and use $\gamma_{a (3+i)} =  \gamma_a \gamma_{3+i}$
recalling $Y^\mu = \bar X^\mu - \bar\theta^{\mu\nu} A_\nu$
and \eq{tilde-gamma},
then the combined transformation 
$\delta = \delta^{(1)} + \delta^{(2)}$
takes the form
\bea
\delta \Psi
 &=& -i \bar\rho^{-1}\,F_{\mu\nu}\tilde\Sigma^{\mu\nu} \epsilon
- \frac 1{\sqrt{\bar\rho}}\,\tilde\gamma^{\mu} \partial_\mu \phi^i  \gamma_{3+i} \epsilon  \nn\\
\delta \phi^i &=& i \bar \epsilon \gamma^{3+i} \Psi, \nn\\
\delta A_\nu &=& 
  -i \sqrt{\bar\rho}\, \tilde g_{\nu\mu}\bar \epsilon \tilde\gamma^\mu \Psi
\eea
where $\tilde\Sigma^{\mu\nu} = \frac i4\,[\tilde\gamma^\mu,\tilde\gamma^\nu]$. The constant factors of $\bar\rho$ can be absorbed
by rescaling the fields.
Noting that $\gamma_{3+i} = \gamma_5\, \Delta_i$
where $\Delta_i$ define the 6-dimensional Euclidean Clifford
algebra,
this indeed amounts to (abelian) $N=4$ supersymmetry 
on $\R^4_\theta$.

In the case of general NC backgrounds,
the SUSY transformation will also act on the metric $\tilde G_{\mu\nu}(y)$.
This can be viewed as a supersymmetric form of emergent gravity, 
which will be worked out elsewhere.
It also means that nontrivial background geometries 
``spontaneously'' break $N=4$ SUSY, as desired.

\section{Discussion and outlook}

In this paper, fermions are studied in the framework of 
emergent noncommutative gravity, as realized through matrix models
of Yang-Mills type. The matrix model strongly suggests a
particular fermionic term in the action, 
corresponding to a specific coupling to
a  background geometry with nontrivial metric $\tilde G_{\mu\nu}$. 
This coupling is similar to the standard coupling of fermions to a 
gravitational background, except that the spin connection vanishes
in the preferred coordinates associated with the matrix model. 

The main result of this paper is that in spite of this unusual
feature, the resulting fermionic action is very reasonable,
and properly describes fermions coupled to emergent gravity. 
In the point particle limit, 
fermions propagate along the appropriate trajectories, albeit with
a different rotation of the spin. At the quantum level, we find an
induced gravitational action which includes the expected
Einstein-Hilbert term with a modified coefficient, 
as well as an additional
term for a scalar density reminiscent of a dilaton. 
There are further terms which vanish for on-shell geometries.
We conclude that the framework of emergent gravity does extend to 
fermions in a reasonable manner, and might well provide
- in a suitable extension - a physically viable theory of gravity.

In a second part of the paper, we compare this induced gravitational 
action 
with the well-known UV/IR mixing in NC gauge theory due to fermions.
Generalizing the results in \cite{Grosse:2008xr} 
for scalar fields, we find 
as expected that the UV/IR mixing can be explained precisely by the
gravitational point of view. 
This also provides a nice understanding for the  fact that 
some UV/IR mixing remains in supersymmetric cases, and only 
disappears for $N=4$ supersymmetry. The reason is that a 
gravitational action is induced even in supersymmetric cases, 
except in $N=4$ SUSY. This in turn leads to the 
conjecture that the gravitational 
constant should be related to the scale of $N=4$ SUSY breaking,
which is quite reasonable. All of these findings suggest that the 
IKKT model
on a noncommutative background 
\cite{Ishibashi:1996xs,Aoki:1999vr,Ishibashi:2000hh,Kitazawa:2005ih} 
should be the most promising candidate for a realistic version of 
emergent gravity.
These issues will be discussed in more detail elsewhere.

\paragraph{Acknowledgments}

We would like to thank H. Grosse for many discussions. 
The work of D.K. was supported by the FWF project P20017, and
the work of H.S. was supported in part by the FWF project P18657 and 
in part by the FWF project P20017.

\section{Appendix A: Computation of $R$}
We quote some identities which appear in the computation
of $R[G]$ and $R[\widetilde{G}]$ in terms of $\theta$-vielbeins:
\bea
\lrk\partial_b G^{bd}\rrk G^{ca}\lrk\partial_a G_{cd}\rrk  &=&
- G^{ac}\lrk\partial_c\thetainv_{ap}\rrk G^{bd}\lrk\partial_d \thetainv_{bq}\rrk g^{pq}
- 2 \lrk\partial_a \theta^{ap}\rrk G^{bc}\lrk\partial_c \thetainv_{bp}\rrk \nn\\
&\quad& 
-\lrk\partial_a \theta^{ap}\rrk\lrk\partial_b \theta^{bq}\rrk g_{pq} 
\\
G^{ab}G^{cd}\partial_b\partial_d G_{ac} 
&=&
2 G^{mp}G^{nq}\thetainv_{ma}\partial_p\partial_q\thetainv_{nb} g^{ab}
+ G^{bc}\lrk\partial_c\thetainv_{ap}\rrk G^{ad}\lrk\partial_d \thetainv_{bq}\rrk g^{pq}
\nn\\
&\quad& 
+ G^{ad}\lrk\partial_d\thetainv_{ap}\rrk G^{bc}\lrk\partial_c \thetainv_{bq}\rrk g^{pq}
\\
G^{mn}G^{pq}\partial_p\partial_q G_{mn} &=&
-2\theta^{mn}G^{pq}\partial_p\partial_q \thetainv_{mn} 
+2 G^{mn}G^{pq}\lrk \partial_p \thetainv_{ma}\rrk\lrk\partial_q\thetainv_{nb}\rrk g^{ab} \\
\lrk\partial_b G^{bd}\rrk\lrk\partial_d \sigma\rrk 
&=&
G^{mn}\lrk\partial_m\sigma\rrk\lrk\partial_n\sigma\rrk
+\theta^{bc}\lrk\partial_b\theta^{da}\rrk\lrk\partial_d\sigma\rrk g_{ca} \\
G^{pq}\lrk\partial_p G^{mn}\rrk\lrk\partial_q G_{mn}\rrk 
&=&
-2 G^{pq}\lrk\partial_p \theta^{mn}\rrk\lrk\partial_q \thetainv_{mn}\rrk\nn\\
&\quad&
-2G^{mn}G^{pq}\lrk\partial_p\thetainv_{ma}\rrk\lrk\partial_q\thetainv_{nb}\rrk g^{ab}\\
G^{np}\lrk\partial_p G^{ac}\rrk\lrk\partial_c G_{na}\rrk
&=&
-2G^{mp}\lrk\partial_p \theta^{nq}\rrk\lrk\partial_n \thetainv_{mq}\rrk 
-\lrk\partial_n\theta^{cq}\rrk\lrk\partial_c\theta^{np}\rrk g_{pq}\nn\\
&\quad&
-G^{nk}\lrk\partial_k \thetainv_{cp}\rrk G^{cl}\lrk\partial_l\thetainv_{nq}\rrk g^{pq}\nn \\
G^{mn}\partial_m\sigma\partial_n\sigma &=& \frac{1}{4} G^{pq}\theta^{mn}\lrk\partial_q\thetainv_{mn}\rrk\theta^{kl}\lrk\partial_q\thetainv{kl}\rrk \nn \\
G^{mn}\partial_m\partial_n \sigma &=& \frac{1}{2}G^{pq}\lrk\partial_p\theta^{mn}\rrk\lrk\partial_q\thetainv_{mn}\rrk 
+\frac{1}{2}\theta^{mn}G^{pq}\partial_p\partial_q \thetainv_{mn}
\eea
Below are some identities which have not appeared in the Ricci scalar 
but appear in the computation of $\tr \cE$:
\bea
G^{mn}\lrk\partial_m G_{np}\rrk g_{ab}\theta^{qa}\lrk\partial_q \theta^{pb}\rrk&=&
-G^{mk}\lrk\partial_k \thetainv_{ma}\rrk G^{nl}\lrk\partial_l\thetainv_{nb}\rrk g^{ab}
-\lrk\partial_m\theta^{ma}\rrk G^{nk}\lrk\partial_k \thetainv_{na}\rrk 
\nn\\
\partial_m \partial_n G^{mn} &=&
2\theta^{ma}\partial_m\partial_n \theta^{nb} g_{ab}
+\lrk\partial_m \theta^{na}\rrk\lrk\partial_n\theta^{mb}\rrk g_{ab}\nn\\
&\quad&
+\lrk\partial_m \theta^{ma}\rrk\lrk\partial_n\theta^{nb}\rrk g_{ab} \nn\\
G_{nk}\theta^{qa}\lrk\partial_q\theta^{nb}\rrk\theta^{lc}\lrk\partial_l\theta^{kd}\rrk g_{ab}g_{cd} &=&
G^{mk}\lrk\partial_k\thetainv_{ma}\rrk G^{nl}\lrk\partial_l\thetainv_{nb}\rrk g^{ab} \nn\\
G_{nk}\theta^{qa}\lrk\partial_q\theta^{nb}\rrk\theta^{lc}\lrk\partial_l\theta^{kd}\rrk g_{ac}g_{bd} &=&
G^{kl}G^{mn}\lrk\partial_k \thetainv_{ma}\rrk\lrk\partial_l \thetainv_{nb}\rrk g^{ab} \nn\\
G_{nk}\theta^{qa}\lrk\partial_q\theta^{nb}\rrk\theta^{lc}\lrk\partial_l\theta^{kd}\rrk g_{ad}g_{bc} &=&
G^{mp}\lrk\partial_p\thetainv_{na}\rrk G^{nq}\lrk\partial_q \thetainv_{mb}\rrk g^{ab} \nn\\
G_{mn}\lrk\partial_p G^{pm}\rrk\lrk\partial_q G^{qn}\rrk &=&
\lrk\partial_p\theta^{pa}\rrk\lrk\partial_q\theta^{qb}\rrk g_{ab}
+2\lrk\partial_p \theta^{pa}\rrk G^{qk}\lrk\partial_k \thetainv_{qa}\rrk \nn\\
&\quad&
+G^{mk}\lrk\partial_k\thetainv_{ma}\rrk G^{nl}\lrk\partial_l\thetainv_{nb}\rrk g^{ab}
\eea

\section{Appendix B: Expressing $R$ and $\tr\, \cE $ in $x$ coordinates}
Let us rewrite the terms which compose $R$ and $\tr \cE$  
in terms of the $\mathfrak{u}(1)$ gauge fields.
We will need ~\cite{Grosse:2008xr}
\bea
e^{\sigma}&=&\lrk \det G^{ab}\rrk^{1/4}=\lrk \det\bar{g}^{ab}\rrk^{1/4}
\left(
1 - \frac 12 \bar{\theta}^{mn}\bar{F}_{mn} + O(\bar{\theta}^2)
\right), \nn\\
G^{ab}&=&\lrk \bar{\theta}^{ac}-\bar{\theta}^{ai}\bar{\theta}^{cj}\bar{F}_{ij}\rrk
       \lrk \bar{\theta}^{bd}-\bar{\theta}^{be}\bar{\theta}^{df}\bar{F}_{ef}\rrk g_{cd},
       \nn\\
\theta^{ab}&=&\bar{\theta}^{ab}-\bar{\theta}^{ac}\bar{\theta}^{bd}\bar{F}_{cd}, \nn\\
\thetainv_{ab}&=&\bar{\theta}^{-1}_{ab}-\bar{F}_{ab}.
\eea
and denote $|\det \bar g_{ab}| \equiv |\bar g_{ab}|$.
For the following terms this gives to $O(A^2)$ in $x$-coordinates
\bea
\int d^4 y \,e^{-\sigma}\,G^{mk}\lrk\partial_k\thetainv_{ma}\rrk G^{nl}\lrk\partial_l\thetainv_{nb}\rrk g^{ab} 
&=&
-\frac{1}{4}\int d^4 x \,|\bar g_{ab}|^{1/4}\bar{\theta}^{mn}\bar
F_{mn}\bar{\partial}^a\bar{\partial}_a\bar{\theta}^{pq}
\bar F_{pq}
\nn\\
\int d^4 y \,e^{-\sigma}\,\lrk\partial_m\theta^{ma}\rrk G^{nk}\lrk\partial_k\thetainv_{na}\rrk
&=&
-\frac{1}{4}\int d^4 x \,
|\bar g_{ab}|^{1/4}\bar{\theta}^{mn}\bar
F_{mn}\bar{\partial}^a\bar{\partial}_a\bar{\theta}^{pq}\bar F_{pq}
\nn\\
\int d^4 y \,e^{-\sigma}\,\lrk\partial_m\theta^{na}\rrk\lrk\partial_n\theta^{mb}\rrk g_{ab}
&=&
-\frac{1}{4}\int d^4 x \,
|\bar g_{ab}|^{1/4}\bar{\theta}^{mn}\bar
F_{mn}\bar{\partial}^a\bar{\partial}_a\bar{\theta}^{pq}\bar F_{pq}
\nn\\
\int d^4 y \,e^{-\sigma}\,\theta^{ma}\partial_m\partial_n\theta^{nb} g_{ab}
&=&
-\int d^4 x \,
|\bar g_{ab}|^{1/4}\Big(
\frac{1}{2}\bar{\partial}^a\bar{\partial}_a\bar{\theta}^{pq}\bar F_{pq}\nn\\
&\quad&
+ \frac{1}{4}\bar{\theta}^{mn}\bar
F_{mn}\bar{\partial}^a\bar{\partial}_a\bar{\theta}^{pq}\bar F_{pq}\Big)
\nn\\
\int d^4 y \,e^{-\sigma}\,G^{pq}\lrk\partial_p\theta^{mn}\rrk\lrk\partial_q\thetainv_{mn}\rrk
&=&
-\frac{1}{2}\int d^4 x \,
|\bar g_{ab}|^{1/4}\bar{\theta}^{mn}\bar
F_{mn}\bar{\partial}^a\bar{\partial}_a\bar{\theta}^{pq}\bar F_{pq}
\nn\\
\int d^4 y \,e^{-\sigma}\,\theta^{mn}G^{pq}\partial_p\partial_q\thetainv_{mn}
&=&
\int d^4 x \,
|\bar g_{ab}|^{1/4}\Big(
-\bar{\partial}^a\bar{\partial}_a\bar{\theta}^{pq}\bar F_{pq} \nn\\
&\quad&
+ \frac{1}{2}\bar{\theta}^{mn}\bar
F_{mn}\bar{\partial}^a\bar{\partial}_a\bar{\theta}^{pq}\bar F_{pq}\Big)
\nn\\
\int d^4 y \,e^{-\sigma}\,G^{kl}G^{mn}\lrk\partial_k\thetainv_{ma}\rrk\lrk\partial_l\thetainv_{nb}\rrk g^{ab}
&=&
\int d^4 x \,
|\bar g_{ab}|^{1/4}\Big(
-\frac{1}{4}\bar{\theta}^{mn}\bar
F_{mn}\bar{\partial}^a\bar{\partial}_a\bar{\theta}^{pq}\bar F_{pq} \nn\\
&\quad&
-\frac{1}{2}\bar g^{ac}\bar g^{bd}\bar F_{ab}\bar{\partial}^2 \bar F_{cd}
\Big)
\nn\\
\int d^4 y \,e^{-\sigma}\,G^{mk}\lrk\partial_k\thetainv_{na}\rrk G^{nl}\lrk\partial_l\thetainv_{mb}\rrk g^{ab}
&=&
-\frac{1}{4}
\int d^4 x \,
|\bar g_{ab}|^{1/4}\bar{\theta}^{mn}\bar
F_{mn}\bar{\partial}^a\bar{\partial}_a\bar{\theta}^{pq}\bar F_{pq} \, 
\eea
where we used~\cite{Grosse:2008xr}
\bea
\int d^4x \,\bar{F}_{mn}\bar{\theta}^{mn}\bar{\partial}^a\bar{\partial}_a \bar{F}_{pq}\bar{\theta}^{pq}
&=&
4\int d^4x \, g^{fh}\bar{F}_{fa}\bar{\partial}^a\bar{\partial}^s \bar{F}_{hs}, \nn\\ 
\int d^4x\,  g^{ab}\bar{F}_{am}\bar{\partial}^m\bar{\partial}^s \bar{F}_{bs}
&=&
\int d^4x \,
\frac 12 \bar{g}^{ah}\bar{g}^{mr}\bar{F}_{hm}\bar{\partial}^2\bar{F}_{ra}
-g^{si}\bar{g}^{ah}\bar{F}_{hs}\bar{\partial}^l\bar{\partial}_l
\bar{F}_{ia} .
\eea

\section{Appendix C: $a^k$ and the spin connection.}

We observe that using the Jacobi equation, the linear term
$a^k$  appearing in $\not \!\! D^2$ \eq{D-square-1}
can be written in the form
\bea
a^k &=&  G^{kl}\partial_l \sigma +\frac{1}{4}\left[\gamma_a,\gamma_b\right]\theta^{lk}\lrk\partial_l\theta^{ab}\rrk - \Gamma^k \nn\\
&=& \frac{1}{4}\left[\gamma_a,\gamma_b\right]\theta^{lk}\lrk\partial_l\theta^{ab}\rrk + \theta^{ka}G^{mn}\lrk\partial_n\thetainv_{ma}\rrk \nn\\
&\stackrel{\mathrm{eom}}{=}& \frac{1}{4}\left[\gamma_a,\gamma_b\right]\theta^{lk}\lrk\partial_l\theta^{ab}\rrk,
\eea
where the last line holds only for on-shell geometries,
using the equations of motion. 
In the standard form of the Dirac operator, 
this linear term would have the form 
\bea
a^k_{\rm comm}&=&\frac 14 \lek\gamma_a,\gamma_b\rek \omega^{a\phantom{l}b}_{\phantom{a}l}G^{lk}-\Gamma^k \nn\\
&=& \frac 14 \left[\gamma_a,\gamma_b\right]\lrk \theta^{kd}g_{dm}\lrk\partial_c\theta^{mb}\rrk g^{ca} 
+ \frac 12 \theta^{lk}\lrk\partial_l \theta^{ab}\rrk \rrk - G^{kl} \partial_l \sigma ,
\eea
where 
\bea
\omega^{a\phantom{k}b}_{\phantom{a}k}&=&-\thetainv_{lc}g^{ca}\nabla_k \theta^{lb} \nn\\
&=&-\thetainv_{lc}g^{ca}\partial_k \theta^{lb}-\thetainv_{lc}g^{ca}\Gamma^{l}_{km} \theta^{mb} \nn\\
&=&\frac{1}{2}\Big(
\theta^{lb}\lrk\partial_k\thetainv_{lc}\rrk g^{ca}-\theta^{na}\lrk\partial_k\thetainv_{nc}\rrk g^{cb} \nn\\
&\quad&-\theta^{mb}\lrk\partial_m\thetainv_{kc}\rrk g^{ca}+\theta^{ma}\lrk\partial_m\thetainv_{kc}\rrk g^{cb} \nn\\
&\quad&-\theta^{mb}\lrk\partial_m\theta^{na}\rrk G_{nk}+\theta^{ma}\lrk\partial_m\theta^{nb}\rrk G_{nk}
\Big) 
=-\omega^{b\phantom{k}a}_{\phantom{a}k}
\eea
is the spin connection, 
using the explicit form of the frame \eq{vielbein}. 
Note that the effect of the spin connection 
is different in our case. Nevertheless, it turns out that 
$\tr\,\cE$ can be rewritten as Ricci scalar $R[\tilde G]$ plus a
term dependent on $\sigma$, as shown in Section \ref{Tr E}.

\end{document}